\documentclass[fleqn,usenatbib]{mnras}

\usepackage{newtxtext,newtxmath}

\usepackage[T1]{fontenc}
\usepackage{ae,aecompl}

\usepackage{graphicx}	
\usepackage{amsmath}	

\newcommand\compsciej{J. Comput. Phys.}  
\newcommand*{\noopsort}[2]{#2#1} 

\title[Anisotropic Conduction in Gas Clouds]{Efficiency of Thermal Conduction in a Magnetised Circumgalactic Medium}

\author[R. Kooij, A. Gr\o nnow, and F. Fraternali]{
Richard Kooij,$^{1}$\thanks{E-mail: grkooij@astro.rug.nl}
Asger Gr\o nnow,$^{1}$
Filippo Fraternali$^{1}$
\\
$^{1}$Kapteyn Astronomical Institute, University of Groningen, Landleven 12, 9747 AD Groningen, The Netherlands\\
}

\date{Accepted 2021 January 11. Received 2021 January 11; in original form 2020 October 8}

\pubyear{2020}

\begin{document}
\label{firstpage}
\pagerange{\pageref{firstpage}--\pageref{lastpage}}
\maketitle

\begin{abstract}
The large temperature difference between cold gas clouds around galaxies and the hot halos that they are moving through suggests that thermal conduction could play an important role in the circumgalactic medium. However, thermal conduction in the presence of a magnetic field is highly anisotropic, being strongly suppressed in the direction perpendicular to the magnetic field lines. This is commonly modelled by using a simple prescription that assumes that thermal conduction is isotropic at a certain efficiency $f < 1$, but its precise value is largely unconstrained. We investigate the efficiency of thermal conduction by comparing the evolution of 3D hydrodynamical (HD) simulations of cold clouds moving through a hot medium,
using artificially suppressed isotropic thermal conduction (with $f$), against 3D magnetohydrodynamical (MHD) simulations with (true) anisotropic thermal conduction. Our main diagnostic is the time evolution of the amount of cold gas in conditions representative of the lower (close to the disc) circumgalactic medium of a Milky Way-like galaxy. We find that in almost every HD and
MHD run, the amount of cold gas increases with time, indicating that hot gas condensation is an important phenomenon that can contribute to gas accretion onto galaxies. For the most realistic orientations of the magnetic field with respect to the cloud motion we find that $f$ is in the range 0.03 -- 0.15. Thermal conduction is thus always highly suppressed, but its effect on the cloud evolution is generally not negligible.
\end{abstract}





\section{Introduction}
Observations have shown that the Circumgalactic Medium (CGM) of galaxies is a complex, multi-phase structure where cold ($T\sim10^4$ K) gas clouds co-exist with an ionised, hot ($T\sim10^6$ K) diffuse halo \citep{Putman_2012, Tumlinson_2017}. The CGM is thought to be maintained by flows of gas away from and towards galaxies, which cause a strong interplay between the hot and cold gas phases. The properties of these hot and cold phases and how they interact with each other are poorly understood. Nevertheless, these interactions are expected to be very important for the process of gas accretion and the feeding of star formation in galaxies.

The idea of a hot halo around the Milky Way arose as a hypothesis by \citet{Spitzer_1956} to provide pressure support to observed gas clouds high above the Galactic disc \citep{Muller_1963}. The existence of such hot halos around galaxies was later theoretically predicted as gas accreted from the intergalactic medium (IGM), that is shock-heated to the virial temperature \citep[][]{White_1978, White_1991, Birnboim_2003}. Hot halos are furthermore consistently formed through accretion in Lambda Cold Dark Matter ($\Lambda$CDM) cosmological simulations \citep[e.g.][]{Fumagalli_2014, Nelson_2016, Fielding_2017}. They are expected to extend to the virial radius ($\sim250$ kpc for the Milky Way), and contain a significant fraction of the baryonic mass of galaxies \citep{Fukugita_2006}, which might be the source of gas for the prolonged star formation observed in typical galactic discs \citep[e.g.][]{Bauermeister_2010}.

The hot phase of the CGM is difficult to detect directly, due to its low X-ray surface brightness \citep[e.g.][]{Bregman_2007}. An X-ray excess around some massive early-type galaxies is known to exist for a long time \citep[e.g.][]{Forman_1979, Forman_1985}. Recently, using the new generation of X-ray telescopes they have also been detected around late-type galaxies \citep[][]{Anderson_2011, Humphrey_2011, Yamasaki_2009, Bogdan_2012, Hodges-Kluck_2014, Anderson_2016}. Evidence in support of a hot galactic halo also comes from absorption line studies towards distant active galactic nuclei (AGN) \citep[e.g.][]{Miller_2013, Stocke_2013}. Additional evidence is obtained indirectly from dwarf satellite galaxies that are devoid of gas, which indicates ram-pressure stripping by a diffuse medium \citep[e.g.][]{Grcevich_2009, Gatto_2013}.

The high-density cold gas phase in the CGM of the Milky Way is detected using H {\small{I}} 21-cm emission, and has revealed a population of extra-planar HVCs \citep[e.g.][]{Muller_1963, Woerden_1991, Wakker_1997, Wakker_2001}. These clouds are characterised by velocities of the order of 100 km/s with respect to the local standard of rest that are inconsistent with the rotation of the Galactic disc. The largest population of extraplanar clouds have deviating velocities with respect to the disc below 100 km/s. These are called Intermediate Velocity Clouds (IVCs). Both HVCs and IVCs in the Milky Way form the extraplanar HI layer \citep{Marasco_2011} which is also observed in external galaxies \citep{Sancisi_2008, Marasco_2019}. HVCs are typically metal-poor, while IVCs have metallicities close to solar \citep{Wakker_2001}. This difference hints at different origins: accretion for the HVCs and feedback for the IVCs. 

It is thought that 
supernovae create hot, ionised super-bubbles that eject gas clouds into the halo regions at velocities on the order of 100 km/s \citep{Maclow_1989}. The ejected gas consists of metal-rich disc material and is likely warm \citep[$\sim10^5$ K,][]{Houck_1990}. As it moves out of the disc it is slowed down by drag forces while cooling and ultimately falls back onto the disc, in a continuous circulation process called the galactic fountain \citep{Shapiro_1976, Fraternali_2006, Grand_2019}. The metal-rich gas clouds from the disc interact with the metal-poor halo material, and gas is stripped from the cloud primarily by Kelvin-Helmholtz instability \citep[KH,][]{Helmholtz_1868, Kelvin_1871}. If the stripped gas mixes efficiently with the hot halo gas it can easily move down the temperature of the latter to where the cooling function peaks ($T\approx$ few $10^5$ K). This causes the cooling time of the mixed gas to become one or more orders of magnitude lower than that of the original hot gas making condensation possible \citep{Marinnaci_2010, Armillotta_2016, Gronke_2020}. The net effect is thus that ejected gas clouds trigger the condensation of hot material from the halo, generally referred to as fountain accretion \citep{Fraternali_2017}. This process can provide the required cold gas to justify the observed star formation rates in galaxies \citep{Fraternali_2008, Marasco_2012, Hobbs_2020}. 

Current knowledge suggests that both the ISM and the CGM of spiral galaxies are magnetised. While much of the precise structure and magnitudes remain unknown, Faraday rotation measures show that the spiral arms of the Milky Way have amplified magnetic fields \citep{Beck_2003, Brown_2007}. The Galactic magnetic field can indeed be of the order of tens of $\mu$G in the spiral arms and the bulge region, while having an average value of $\approx3\mu$G \citep{Beck_2009, Beck_2016}. 

Magnetic fields are also observed outside the discs of spiral galaxies, in the halo region \citep[e.g.][]{Ekers_1977, Sancisi_1979, Irwin_2012}. It remains however very hard to infer the magnetic field orientation and strength in the halo of the Milky Way. Based on combined Wilkinson Microwave Anisotropy Probe 5 \citep[WMAP5, ][]{Komatsu_2009} polarisation data and Faraday rotation measures, it was shown by \citet{Jansson_2009} that most Galactic field models predict magnetic fields that are inconsistent with observations. The halo region likely requires its own component in a galactic field model, rather than being a simple extension of the disc field. Observations of external galaxies show that an additional out-of-plane component is also required \citep{Beck_2009, Krause_2009}. Turbulent gas flows in galactic halos also give rise to an amplified small scale random magnetic field component \citep[e.g.][]{Beck_2016}.

Magnetic fields can suppress the formation of KH instabilities \citep{Chandrasekhar_1961}. This also occurs in cloud-wind simulations \citep{Maclow_1994, Jones_1996} and generally extends the lifetimes of the clouds. Additionally, under typical CGM conditions the magnetic field drapes around the cloud leading to locally amplified magnetic fields that can become dynamically important and shield the cloud from destruction \citep[e.g.][]{Dursi_2008, Banda-Barragan_2016, Gronnow_2017, Gronnow_2018}. The magnetic draping effect is also seen in simulations that include random magnetic fields \citep[e.g.][]{Asai_2007, Sparre_2020}. The field lines in these simulations are draped along the cloud and are mostly ordered. Turbulent magnetic fields thus become mostly ordered locally around IVCs and HVCs. This is supported by recent highly detailed Faraday rotation measures near the Smith HVC, that also show signs of magnetic draping \citep{Betti_2019}. 

 The primary effects on the evolution of fast moving cold clouds in hot halos are expected to be from radiative cooling, thermal conduction, magnetic fields, and in some cases self-gravity \citep{Li_2020}. However, the inclusion of radiative cooling, magnetic fields and thermal conduction in simulations is complicated and has only recently been investigated \citep[][but see also \citealp{Orlando_2008} for the related cloud-shock case]{Li_2020, Liang_2020}. Thermal conduction is a diffusive heat exchange process that takes place in the presence of strong temperature gradients \citep{Spitzer_1962}. It is therefore expected to play a fundamental role in the evolution of the CGM due to the large temperature difference between the cold cloud and the hot halo, and has been included by several authors \citep[e.g.][]{Vieser_2007, Bruggen_2016, Armillotta_2016, Armillotta_2017, Li_2020, Liang_2020}. The thermal conduction heat flux at high temperatures is dominated by the contributions of free electrons, that preferentially move along magnetic field lines. The heat flux in the presence of magnetic fields is thus highly anisotropic, being strongly suppressed perpendicularly to field lines (see Section~\ref{sec:numerical_simulations} for a detailed description). However, running full MHD simulations with thermal conduction can be very costly and several authors have resorted to include thermal conduction in hydrodynamical simulations in the form of an isotropic conduction, while mimicking the effect of magnetic fields using a suppression factor. This later point is crucial because even in a weak and dynamically unimportant magnetic field, thermal conduction perpendicular to field lines is still dramatically suppressed. 
 \citet{Bruggen_2016} shows indeed that isotropic unsuppressed thermal conduction evaporates clouds on timescales that are too short as compared to observational constraints. Thermal conduction at Spitzer value is thus clearly unphysical. For this reason, thermal conduction is generally assumed to operate at a certain efficiency $f<1$. \citet{Narayan_2001} show that the efficiency of thermal conduction in a tangled magnetic field can be reduced to $20\%$ ($f=0.20$) of the classical Spitzer value. Other authors find that thermal conduction is much less efficient at $0.1-1\%$  ($f\approx0.001-0.01$) of the Spitzer value \citep{Chandran_1998}. As we previously mentioned, the fast moving clouds will experience a draped magnetic field instead of a tangled magnetic field. It is therefore unclear how efficient thermal conduction is in the CGM, and thus also its importance. Typical values that are used in literature are $10\%$ \citep[$f=0.1$, ][]{Armillotta_2016, Armillotta_2017}. 

In this paper we investigate the approximation of the efficiency $f$-factor by running a large suite of fully 3D HD and MHD simulations of cloud-wind systems representative of the conditions in the lower (close to the galaxy) CGM of Milky Way-like galaxies. Our main goal is to compare the evolution between HD and MHD simulations, focusing specifically on the evolution of the cold gas mass. Since we include radiative cooling in all of our simulations, the net amount of cold gas will generally increase \citep[condensation, e.g.][]{Marinnaci_2010}, as opposed to decrease (evaporation). However, given that both thermal conduction \citep{Armillotta_2016}, and magnetic fields \citep{Gronnow_2018} have been shown to suppress condensation, the quantification of this reduction is a primary goal in this paper.

This paper is organised as follows. In Section~\ref{sec:Methods} we describe the code used for our simulations and the numerical setup. In Section~\ref{sec:Results} we show our results, which are discussed in Section~\ref{sec:discussion}. We conclude in Section~\ref{sec:conclusions}.

\section{Methods} \label{sec:Methods}
\subsection{Simulation Code}
We use \textsc{pluto} version 4.3 \citep{Mignone_2007_pluto, Mignone_2011_amr} to solve the system of equations for hydrodynamical (HD) and magnetohydrodynamical (MHD) flow in 3 spatial dimensions. \textsc{pluto} is a Eulerian Godunov-type \citep{Godunov_1959} code. For both HD and MHD flows we close the system of equations with an ideal equation of state. For the internal energy we assume that both the cloud and hot halo are monatomic and accordingly set the adiabatic index $\gamma = 5/3$. In MHD runs we approximately enforce the solenoidal constraint ($\nabla \cdot \bmath{B} = 0$) using the hyperbolic divergence cleaning scheme by \citet{divclean_Dedner_2002}, implemented in \textsc{pluto} by \citet{divclean_mignone_2010}. For the flux computations there is a trade-off to be made between accuracy and numerical stability, as more accurate Riemann solvers are generally also less stable. As a compromise, we employ the approximate Riemann solver HLLC \citep{HLLC_Toro_1994} for HD simulations and HLLD \citep{HLLD_Miyoshi_2005} for MHD simulations. Note that the HLLD Riemann solver is merely an MHD extension of the HLLC Riemann solver, and does not introduce artificial differences between HD and MHD simulations. We use second-order Runge-Kutta time integration and linear (2nd order) reconstruction for all simulations.

In order to run our simulations at sufficient spatial resolutions while keeping computational costs feasible, we employ adaptive mesh refinement (AMR). We use the gradient of the density as refinement variable, and to ensure numerical stability we set the Courant-Friedrichs-Lewy \citep[CFL,][]{Courant_1928} number to $C_\alpha$ = 0.3 for all simulations. We start our simulations from a base grid of $N_x \times N_y \times N_z = 60\times20\times20$ on a physical grid of $L_x\times L_y\times L_z = 6\times2\times2$ kpc. For each level of refinement we increase the resolution in every dimension by a factor 2. In the fiducial setup we refine 5 times to an equivalent resolution of $1920\times640\times640$. In this way, there are 32 cells per cloud radius ($\mathcal{R}_{32}$), which gives a spatial resolution of $\sim$3 pc/cell at the highest refinement level.

\subsection{Numerical simulations}
\label{sec:numerical_simulations}
We initialise our simulations according to the typical "cloud-wind" problem, and follow the general setup as described in \citet{Gronnow_2018}. In particular, we assume the same steep, smooth density profile given by
\begin{equation}
    \rho(r) = \rho_{\rm hot} + \frac{1}{2}(\rho_{\rm cloud} - \rho_{\rm hot})\times\Bigl(1 - \tanh\Bigl[s\Bigl(\frac{r}{r_0}-1\Bigr)\Bigr]\Bigr),
\end{equation}
where $\rho_{\rm hot}$ is the density of the hot halo, $\rho_{\rm cloud}$ is the density of the cloud, $r_0$ is the cloud radius, and $s=10$ is the steepness parameter. Thus, the cloud radius $r_0$ for this density profile is defined to be at the halfway point between the central cloud density and the halo density. Turbulent velocities with Mach-number $\mathcal{M}\approx1$ are added to the cloud to make the simulations more realistic \citep{Marasco_2019}, and to reduce artifacts arising from symmetry. Similar to \cite{Armillotta_2016}, we pick random values from a Gaussian distribution of velocities with a dispersion of 10 km/s, and a mean of 0 km/s. These velocities are then assigned to the region consisting of only cloud material i.e. $r<r_{\rm 0}$, where $r_0$ is the cloud radius. 

\begin{table*}
    \centering
    \parbox{10.4cm}{
    \caption{The cloud and hot halo parameters for our fiducial setup. }
    }
    \label{tab:default_parameters}
    \begin{tabular}{p{1cm}p{1cm}p{1cm}p{1cm}p{1cm}p{1cm}p{1cm}p{0.3cm}}
        \hline
         $v_\text{rel}$\ \ \ \ (km s$^{-1}$) & $T_\text{cloud}$ (K) & $T_\text{hot}$\ \ \ \ \ (K) & $Z_{\text{cloud}}$ ($Z_{\odot}$) & $Z_{\text{hot}}$ ($Z_{\odot}$) & $n_{\rm cloud}$ (cm$^{-3}$) & $n_\text{hot}$ (cm$^{-3}$) & $r_{0}$\ \ \ \ \ \ \ \ \   (pc)   \\
         \hline
          75 & 10$^4$ & 2$\times$10$^6$ & 1  &  0.1  & 0.2 &  $10^{-3}$  &  100  \\
         \hline
    \end{tabular}
    \parbox{10.4cm}{
    \textit{Notes.} The parameters are the relative velocity between cloud and halo ($v_\text{rel}$), cloud temperature ($T_\text{cloud}$), halo temperature ($T_\text{hot}$), cloud/halo metallicity ($Z_\text{cloud/hot}$), halo number density ($n_\text{hot}$), and cloud radius ($r_0$). Note that the pressure equilibrium between cloud and halo makes the cloud number density a fixed value.
    }
    
\end{table*}

We include radiative cooling in every simulation using the tabulated cooling module, based on the collisional ionization equilibrium cooling tables of \citet{Sutherland_1993}. The variation in mean molecular weight $\mu$ with temperature $T$ is accounted for, and we interpolate for the variation in metallicity ($Z$) with $T$ between 3 tables with metallicities 0.1$Z_\odot$, 0.3$Z_\odot$, $Z_\odot$, similarly to \citet{Marinacci_2011} and \citet{Gronnow_2018phd}. We keep track of the metallicity using a passive scalar $C$ that does not affect the flow, which we advect conservatively as
\begin{equation}\label{eqn:passive_scalar}
    \frac{\partial (\rho C)}{\partial t} + \nabla\cdot(\rho C\bmath{v}) = 0.
\end{equation}
We initialise the passive scalar $C$ in simulations at a cloud metallicity ($Z_\text{cloud}$) for $r<r_\text{0}$, and at a hot halo metallicity $Z_\text{hot}$ elsewhere. We finally assume a cooling floor at $T=10^4$ K below which no cooling occurs. This assumption is justified as the cooling rate drops drastically below $10^4$ K, and most of the CGM "cold" material is observed at these temperatures \citep{Putman_2012, Werk_2013}.

Thermal conduction is included in our simulations using the readily available thermal conduction module in \textsc{pluto}, but with the following modifications. We assume that thermal conduction happens exclusively through free electrons, and we account for this by multiplying the thermal conduction heat flux by the ionisation fraction $x_{\rm i}$. The ionisation fraction depends on the temperature, which we calculate using the same procedure as described for radiative cooling.

The full equation for the thermal conduction heat flux becomes
\begin{equation} \label{eqn:conduction}
    \bmath{q} = -f\frac{\kappa_{\rm Sp}\nabla T}{1 + \sigma}x_{\rm i},
\end{equation}
where $\bmath{q}$ is the heat flux, $x_{\rm i}$ is the ionisation fraction, $\nabla T$ is the gradient of the temperature, and $\kappa_\text{Sp}$ is the classical Spitzer \citep{Spitzer_1962} conductivity given by
\begin{equation}
    \kappa_\text{Sp}= 1.84\times10^{-5}\frac{T_{\rm e}^{5/2}}{\text{ln}\ \Psi}\ \ \ \  \text{ergs/s/K/cm},
\end{equation}
where $\nabla T$ is the gradient of the temperature, $T_{\rm e}$ is the electron temperature which we assume to be equal to the fluid temperature, and ln $\Psi$ is the Coulomb logarithm given by
\begin{equation}
    \text{ln}\ \Psi = 29.7 + \text{ln} \frac{1}{\sqrt{n_{\rm e}/\text{cm}^{-3}}}\frac{T_{\rm e}}{10^6 \text{K}},
\end{equation}
where $n_{\rm e}$ is the electron number density. 
When the mean free path of electrons becomes similar to or greater than the temperature scale length, the classical Spitzer flux is no longer accurate and the heat flux is said to be saturated. We account for saturated heat flux according to the description as given by \citet{Cowie_1977}, and we set
\begin{equation}
    \bmath{q}_\text{sat} = 5\phi_\text{sat}\rho c_{\rm s}^3,
\end{equation}
where $\bmath{q}_\text{sat}$ is the saturated heat flux, $\rho$ is the gas density, $c_{\rm s}$ is the local sound speed, and $\phi_\text{sat}$ is a factor of order unity that accounts for uncertainties regarding the flux-limited treatment. Following the same arguments as presented in \citet{Armillotta_2016} we set $\phi_\text{sat} = \sqrt{f}$. To ensure a smooth transition between the classical and saturated regimes the factor $1 + \sigma$ is included in Eq.~\ref{eqn:conduction}, defined as
\begin{equation}
    \sigma = \frac{\kappa_\text{Sp}{\lVert}\nabla T{\rVert}}{\bmath{q}_\text{sat}},
\end{equation}
where ${\lVert}\nabla T{\rVert}$ is the magnitude of the temperature gradient. Finally, the efficiency factor $f$ is a dimensionless parameter between 0 and 1 that is typically used to approximate the suppression due to magnetic fields as discussed above, and is the focus of this work.

In a subset of our simulations we include magnetic fields. We vary between weak (0.1 $\mu$G) and strong (1.0 $\mu$G) magnetic fields, as appropriate for the lower part of the CGM \citep{Gronnow_2018}. For each field strength we run the simulations for orientations of the magnetic field parallel ($B_\parallel$) or perpendicular ($B_\bot$) to the relative velocity. In simulations where we include both thermal conduction and a magnetic field, we calculate the anisotropic thermal conduction heat flux, which is natively included in \textsc{pluto}. In this case the heat flux is split up into parallel and perpendicular components with respect to the direction of the magnetic field as follows \citep{Spitzer_1962, Braginski_1965}.
\begin{equation}\label{eq:anisotopric_conduction}
    \bmath{q}_{\rm class} = \kappa_{\parallel}\bmath{\hat{b}}(\bmath{\hat{b}}\cdot{\nabla} T) + \kappa_{\bot}[{\nabla} T - \bmath{\hat{b}}(\bmath{\hat{b}}\cdot{\nabla} T)],
\end{equation}
where $\bmath{q}_{\rm class}$ is the classical conduction heat flux, $\bmath{\hat{b}} = \bmath{B}/|\bmath{B}|$ is a unit vector denoting the direction of the magnetic field, and $\kappa_\bot$ and $\kappa_\parallel$ are the perpendicular and parallel components of the conductivity, where $\kappa_\parallel=\kappa_{\rm Sp}$. The perpendicular conductivity is given by
\begin{equation}
    \kappa_{\bot} = \frac{8\sqrt{\pi m_{\rm H}k_{\rm B}}n_{\rm H}^2e^2c^2\ln  \Psi}{3|\bmath{B}^2|\sqrt{T}},
\end{equation}
where $m_{\rm H}$ is the mass of hydrogen, $k_{\rm B}$ is Boltzmann's constant, $n_{\rm H}$ is the hydrogen number density, and $c$ is the speed of light in vacuum. For the parameters used in our analysis, $\frac{\kappa_\bot}{\kappa_\parallel} \lesssim 10^{-9}$. The conductivity perpendicular to field lines is thus much smaller than that parallel to the field lines, and we set $\kappa_\bot = 0$.

 To assess the effect of isotropic thermal conduction, we perform HD simulations with either only radiative cooling, or radiative cooling and isotropic thermal conduction at several efficiencies by varying $f$, as listed in Table~\ref{tab:variation_parameters}. To examine the effect of anisotropic thermal conduction in the presence of magnetic fields, we run MHD simulations at the field strengths and orientations as mentioned above. In this case, we set anisotropic thermal conduction either off, or at full Spitzer efficiency ($f = 1$). Our main goal is to compare the evolution between isotropic thermal conduction runs (no magnetic field) and anisotropic thermal conduction runs (with magnetic fields).
 
 Additionally, we explore several different cloud and halo parameters. We investigate a lower cloud metallicity, a higher relative velocity, and lower cloud and halo densities. For every different set of parameters, we perform the same suite of HD and MHD simulations as described above (Table~\ref{tab:variation_parameters}). 

\begin{table}

    \caption{The varied simulation parameters.}
    \label{tab:variation_parameters}
    \begin{tabular}{ |p{1.5cm}|p{1.5cm}|p{4cm}| } 
        \multicolumn{3}{|c|}{} \\
        \hline
         Parameter & Setup & Values considered  \\
         \hline
         $f$           &  HD &  0.01, 0.05, 0.1, 0.15, 0.2 \\
         $v_\text{rel}$ $^a$   &  HD,MHD       &   $\mathbf{75}$, 150 [km/s]  \\
         $Z_\text{cloud}$    &  HD,MHD        &   0.1, $\mathbf{1.0}$ [$Z_\odot$]  \\
         $n_\text{cloud}$,$n_\text{hot}$ $^b$   &  HD,MHD         &   ($\mathbf{0.2}, \mathbf{10^{-3}}$), (0.1, 5$\times10^{-4}$) [cm$^{-3}$] \\
         $|\bmath{B_0}|$ $^c$   &  MHD         &   0.1, 1.0 [$\mu$G]  \\
         $B_\text{orientation}$   &  MHD         &   $B_\bot$, $B_\parallel$, $B_{\rm ob}$  \\
         
         \hline
    \end{tabular}
    
    \footnotesize{\textit{Notes.} The 'Setup' column describes for which module(s) we perform the variation. The values used in the fiducial simulation setup are highlighted in bold.
    $^a$This corresponds to Mach numbers of the hot gas $\mathcal{M}\approx0.35$ and 0.70 for relative velocities 75 and 150 km/s, respectively. $^b$We assume pressure equilibrium between the cloud and the hot halo, thus the density contrast between cloud and halo is in both cases $\chi = \frac{n_\text{cloud}}{n_\text{hot}}=200$. $^c$The plasma-$\beta$ parameter, defined as the ratio between the thermal pressure of the gas over the magnetic pressure, is for the higher densities $\beta$=7 and 700 for field strengths 1 and 0.1 $\mu$G, respectively. For the lower density setup, the plasma-$\beta$ values are a factor 2 smaller.}\\
\end{table}

\subsection{Analysis of the simulations} \label{sec:analysis}
To follow the evolution of simulations with different parameter setups equally, we denote time in units of cloud-crushing time $t_\text{cc}$ \citep{Jones_1996},
\begin{equation}
    t_\text{cc} = \sqrt{\chi}\frac{2r_{\rm 0}}{v_\text{rel}},
\end{equation}
where $\chi = \frac{n_\text{cloud}}{n_\text{hot}}$ is the density contrast, $r_{\rm 0}$ is the radius of the cloud, and $v_\text{rel}$ is the relative velocity between the cloud and the halo. For our fiducial setup, $t_\text{cc} \approx 18$ Myr.

We diagnose our data using volume averaged, mass weighted quantities as described in \citet{Klein_1994} and \citet{Shin_2008}. We focus on the evolution of the amount of cold gas in the simulation domain according to
\begin{equation}\label{eqn:coldgas}
    \Delta \text{M}_\text{cold} = \frac{\int \rho_\text{cold} dV}{M_\text{cloud, $t$=2 Myrs}},
\end{equation}
where, similar to \citet{Armillotta_2016}, we denote gas as "cold" if $T < 2\times10^4$ K, and we normalise by the cloud mass at $t=2$ Myrs $M_\text{cloud, $t$=2 Myrs}$. We use the cloud mass after 2 Myrs as opposed to the initial cloud mass since the transition region between cloud and halo consists of intermediate temperature ($T\approx10^5$ K) material that quickly cools down to $10^4$ K which leads to an early jump in the cold gas mass that is not due to condensation.

We furthermore calculate the mass of mixed gas similar to \citet{Xu_1995}. We define a passive scalar $C$ over the simulation volume, where we set $C=1$ inside the cloud, $C=0$ elsewhere. This is advected using Eqn.~\ref{eqn:passive_scalar}, and we define gas to be 'mixed' if $0.1 < C < 0.9$. 
\section{Results}\label{sec:Results}
\subsection{Fiducial simulation setup}
Our fiducial simulation setup focusses on cool clouds travelling through the lower CGM of a Milky Way-like galaxy. We list the parameters used in our fiducial simulation setup in Table~\ref{tab:default_parameters}. We assume parameters for a typical intermediate velocity cloud, that are consistent with earlier work \citep[e.g.][]{Marinnaci_2010, Armillotta_2016, Gronnow_2018}. We furthermore assume a halo temperature typical of the Milky Way \citep[e.g.][]{Bregman_2007}, and a halo metallicity ($Z_{\rm hot}$) as estimated in Milky Way-like galaxies \citep[e.g.][]{Bogdan_2017, Hodges-Kluck_2018}. The clouds considered in this work exceed the critical "mass growth" radius of \citet{Gronke_2018, Gronke_2020} and are thus in the mass growth regime, this agrees with the fact that we find condensation.
\citet{Li_2020} propose a qualitatively different threshold for which the column density of the cloud can be used to distinguish between cloud destruction and mass growth.
In our fiducial setup the clouds are {\em just} within their “classical cloud destruction” regime and thus the discrepancy with this criterion does not seem significant; in this respect see also \citet{Kanjilal_2020}. 
In our simulations with a lower cloud column density (Section~\ref{sec:lowrho}) that more clearly belong to the cloud destruction regime of \citet{Li_2020} we do not see a discrepancy, as they indeed do not show condensation.

\subsubsection{HD simulations}
\begin{figure*}
    \centering
    \includegraphics[width=\textwidth]{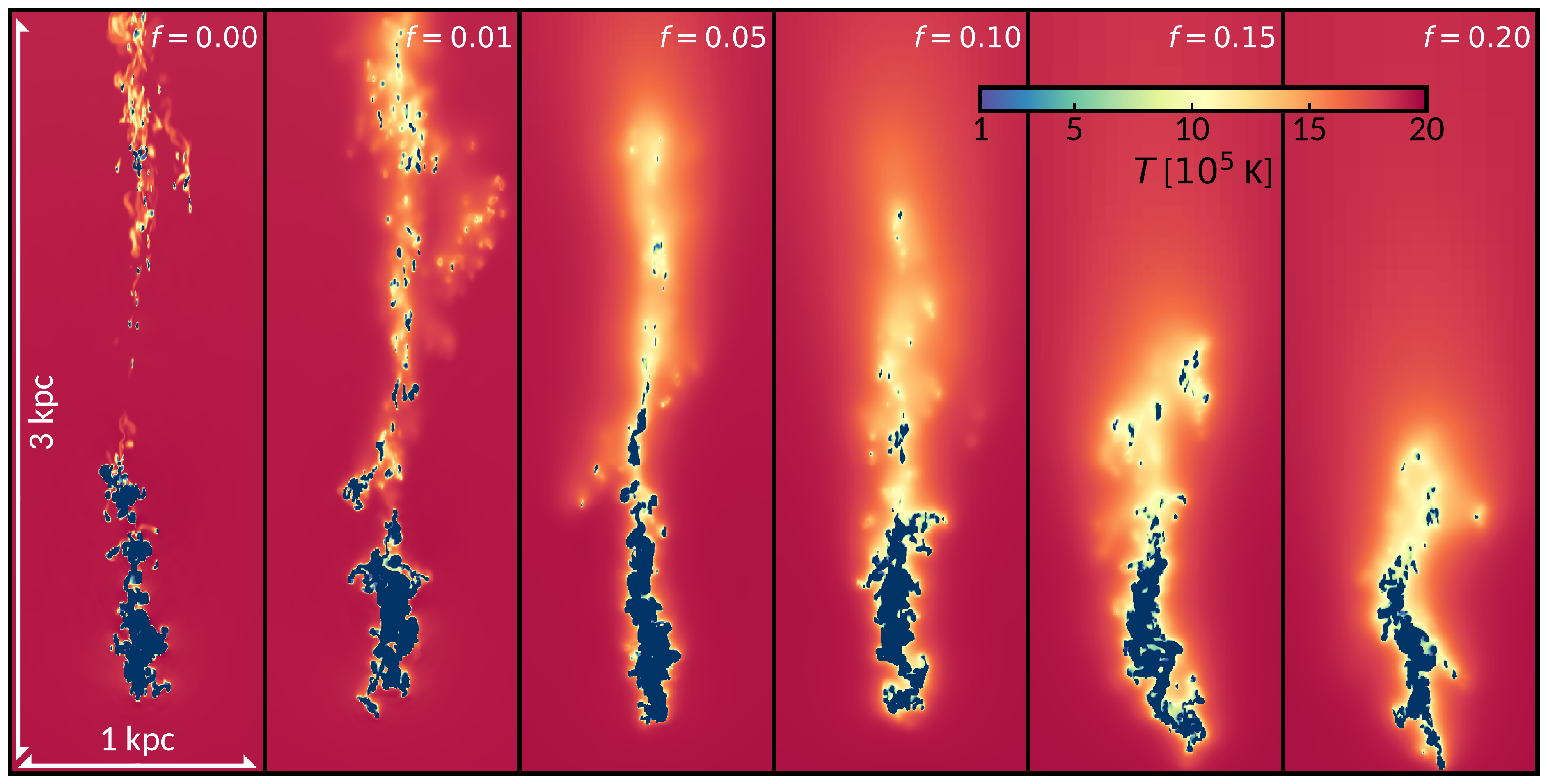}
    \caption{Slices through part of the simulation domain of the gas temperature at $t=60$ Myr for the fiducial simulation setup (Table \ref{tab:default_parameters}). We vary the efficiency of isotropic thermal conduction with the $f$-factor from $f=0.0$ (no thermal conduction) to $f=0.20$ (strong thermal conduction) of the Spitzer value. All gas below $2\times10^4$ K has been coloured dark blue.}
    \label{fig:solar_CTC}
\end{figure*}

\begin{figure}
    \includegraphics[width=0.5\textwidth]{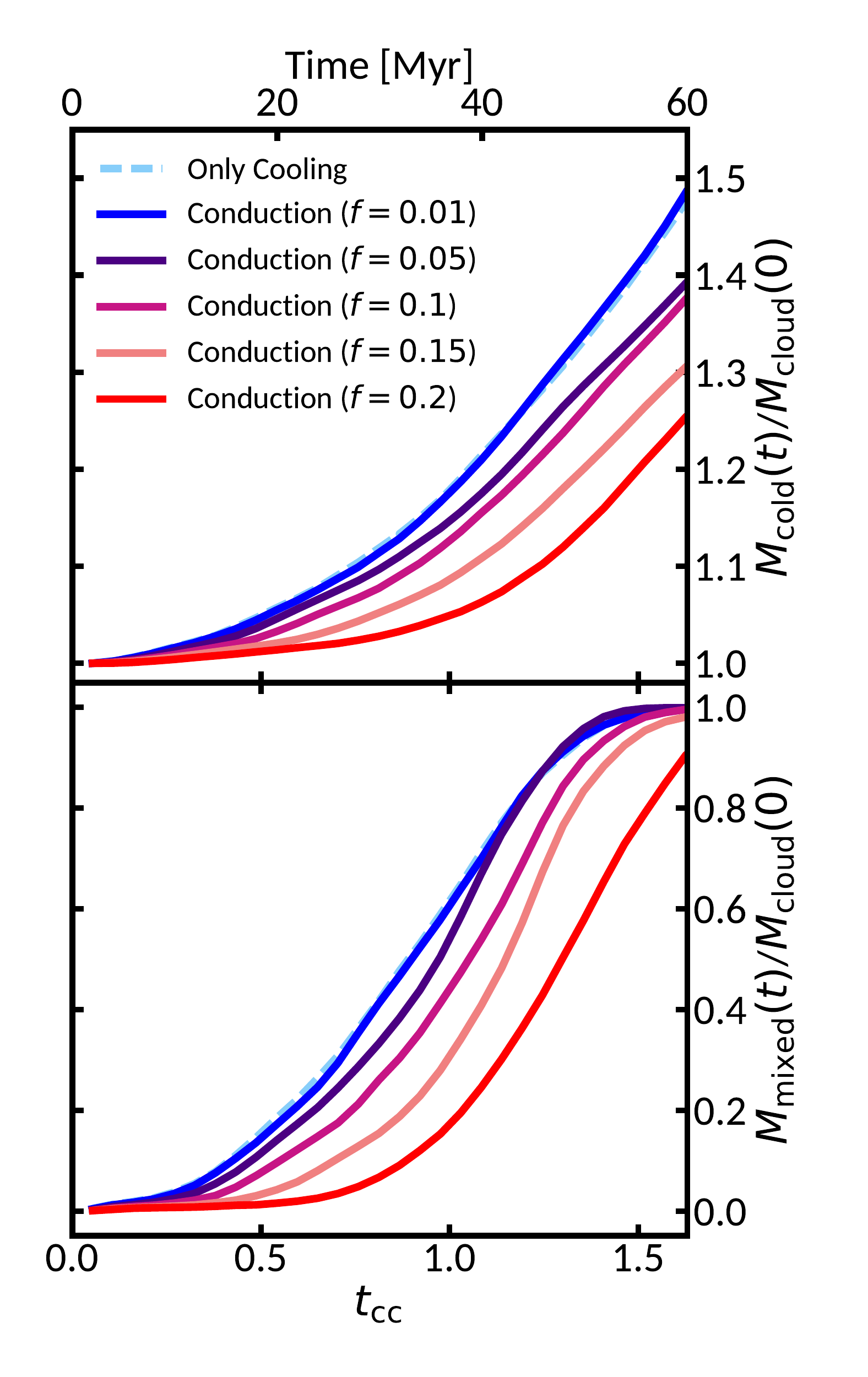}
    \caption{We show the cold gas mass evolution (top) for our fiducial simulation setup (Table~\ref{tab:default_parameters}) for HD runs. We vary only the efficiency of isotropic thermal conduction through the parameter $f$. On the bottom, we show the mixing fraction for the same simulations, where the colour coding is the same as for the top panel.}
    \label{fig:solar_condensation}
\end{figure}

We investigate the effects of isotropic thermal conduction in 3D HD simulations by varying the efficiency of conduction through the $f$-factor. We show slices of the temperature distribution through part of the simulation domain after $60$ Myrs in Fig.~\ref{fig:solar_CTC}. Similar to previous work, e.g. \citet{Bruggen_2016, Armillotta_2016}, we find that thermal conduction decreases the size of the wake through the evaporation of stripped cloudlets. The amount of cold gas in the simulation domain increases (i.e. there is condensation) in all cases, which is displayed in Fig.~\ref{fig:solar_condensation}, top panel. However, the amount of condensation decreases with stronger (higher value of $f$) thermal conduction. Additionally, we show the mixed gas evolution in Fig.~\ref{fig:solar_condensation}, bottom panel. All runs except $f=0.2$ are fully mixed ($M_\text{mixed}/M_\text{cloud, $t$=2 Myrs} = 1$) after $60$ Myrs. This noticeable decrease in mixing efficiency for the $f=0.2$ run shows that thermal conduction can have an additional effect, which we expand on later in this section. 

According to the work by \citet{Begelman_1990}, the thermal instability is suppressed on scales smaller than the Field length $\lambda_{\rm F}$, named after the work on thermal instability by \citet{Field_1965}:
\begin{equation}\label{eqn:Field}
    \lambda_{\rm F} = \sqrt{\frac{f\kappa_\text{Sp}T_\text{hot}}{n^2_\text{cold}\Lambda(T_\text{cold})}},
\end{equation}
where $n_{\rm{cold}}$ is the number density of cold (in our case $\sim10^4$ K) material, and $\Lambda(T_{\rm{cold}})$ is the cooling rate at this temperature\footnote{Since we implement a cooling floor at $T=10^4$ K, in principle there is no cooling at this temperature. Instead, we consider the cooling rate slightly above $10^4 K$ and take $\Lambda(T_{\rm cold}) = 10^{-24}$ ergs cm$^3$/s.}. For our fiducial simulations the Field length varies in the range $\sim10-60$ pc for $f=0.01-0.20$, respectively. Hence, for weak thermal conduction ($f=0.01$) the Field length is slightly below the size of stripped cloudlets ($\sim10$ pc) and they do not evaporate. However, stronger thermal conduction increases the Field length and thus quickly dominates over cooling making the cloudlets evaporate. We also see the formation of a smooth, intermediate temperature wake. Stripped cloudlets inside this intermediate temperature wake are more stable against evaporation since $T_{\rm hot}$ is smaller, which decreases the Field length. It is for this reason that some cloudlets are seen to survive in the wake even for strong ($f=0.2$) thermal conduction. Note also that in the strong thermal conduction cases ($f=0.15-0.2$), the Field length is $\approx 50$ pc, which evidently has a strong effect on the evolution of the cloud itself. We argue that in this case, thermal conduction can form and maintain a smooth temperature gradient between the head of the cloud and the hot halo. This subsequently creates a less steep density gradient that can suppress the formation of KH-instabilities \citep{Vieser_2007}, which extends the lifetime of the cloud \citep{Armillotta_2017}.  

Therefore the primary effect of thermal conduction is the evaporation of stripped cloudlets. However, in cases where thermal conduction is very strong, a secondary effect can become dominant that suppresses KH instabilities and delays mixing and condensation. 

\begin{figure*}
    \centering
    \includegraphics[width=\textwidth]{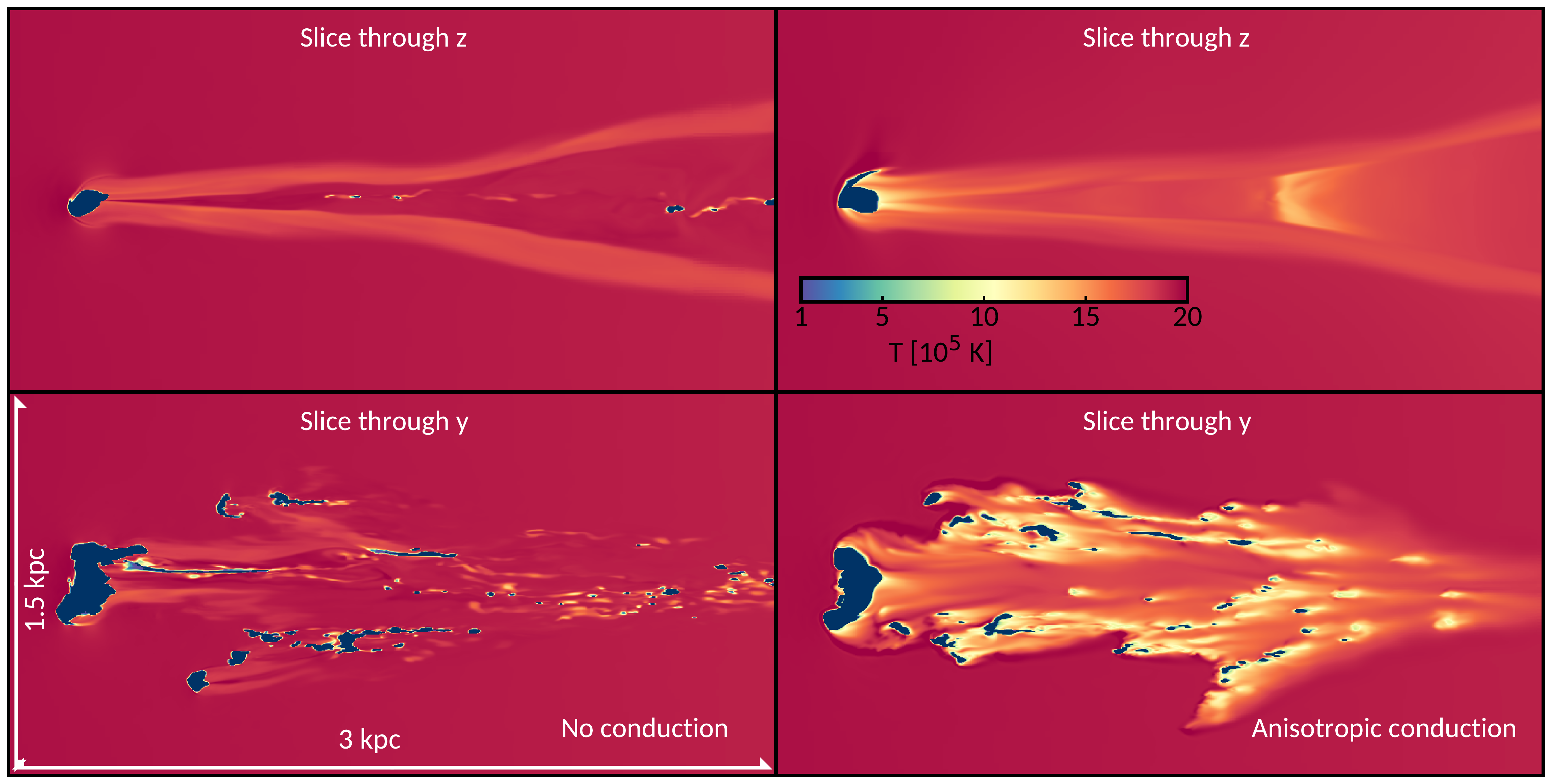}
    \caption{Slices through part of the simulation domain of the temperature in the weak, perpendicular magnetic field runs for our fiducial simulation setup (Table~\ref{tab:default_parameters}). To illustrate the anisotropic morphology we show slices through the y- and z-axes. The magnetic field is initially parallel to the y-axis. We show the simulation without thermal conduction on the left, and the simulation with anisotropic conduction on the right. Same as for Fig.~\ref{fig:solar_CTC}, all gas below $2\times10^4$ K has been coloured dark blue.}
    \label{fig:mag_field_temperatures}
\end{figure*}

\begin{figure*}
    \centering
    \includegraphics[width=\textwidth]{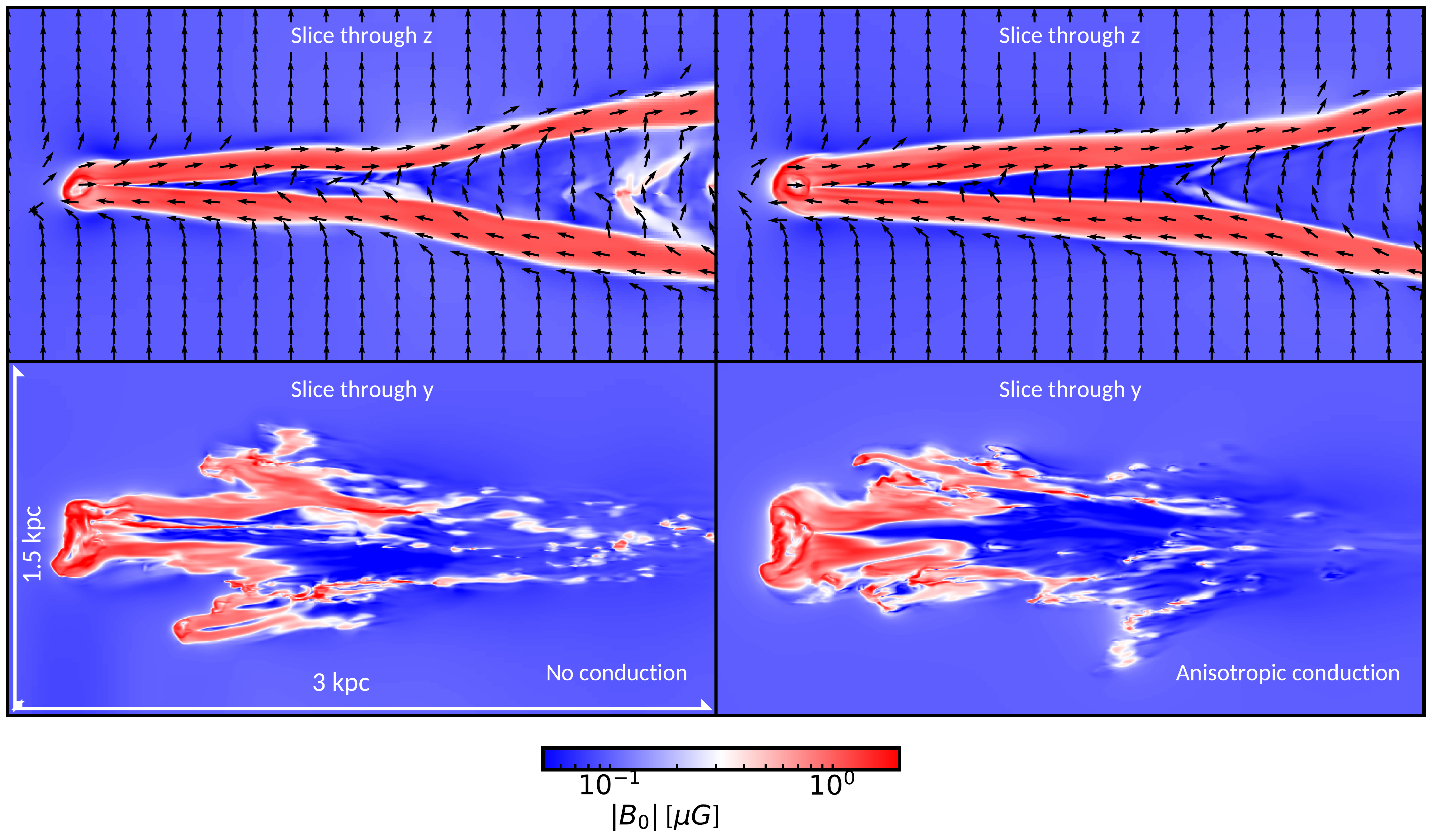}
    \caption{The same simulations as shown in Fig.~\ref{fig:mag_field_temperatures}, but here we show slices of the magnetic field strength (colours) and orientation (arrows). We overlay the normalised magnetic field vectors on the top panels. We do not include the field vectors in the bottom panels since the field is mostly oriented into the paper. }
    \label{fig:mag_field_perp}
\end{figure*}
\subsubsection{MHD simulations}
We perform MHD simulations with and without anisotropic thermal conduction, and we show temperature slices for a weak (0.1 $\mu$G), magnetic field perpendicular to the motion of the cloud ($B_{x} = B_{z} = 0$, and $B_{y}= B_0$) in Fig.~\ref{fig:mag_field_temperatures}. The dynamical evolution of the system is affected by the magnetic field even in the weak field setup. By a magnetic draping effect \citep{Dursi_2008, Gronnow_2018}, field lines are swept up and stretched by the cloud leading to significant field amplification as we show in Fig.~\ref{fig:mag_field_perp}. Magnetic pressure becomes dynamically important and squeezes cloud material along one axis (see e.g. the top panels of Fig.~\ref{fig:mag_field_temperatures} and \ref{fig:mag_field_perp}), which causes the cloud to expand along the axis perpendicular to it (bottom panels of Fig.~\ref{fig:mag_field_temperatures} and \ref{fig:mag_field_perp}). The overall morphology of the cloud gas is not strongly affected by the inclusion of anisotropic thermal conduction in this case. However, we note that thermal conduction creates a wake with temperatures of $T \approx1-1.5\times10^6$ K. Due to the magnetic draping effect the field lines are wrapped around the head of the cloud and cloudlets. Since thermal conduction is only efficient along the field lines, a steep temperature gradient between cloud and hot halo can be maintained for at least $60$ Myrs. Additionally, we notice that small stripped cloudlets can survive even with thermal conduction included, which is consistent with the findings of \citet{Liang_2020} in their 2D MHD simulations.

\begin{figure*}
    \centering
    \includegraphics[width=1.0\textwidth]{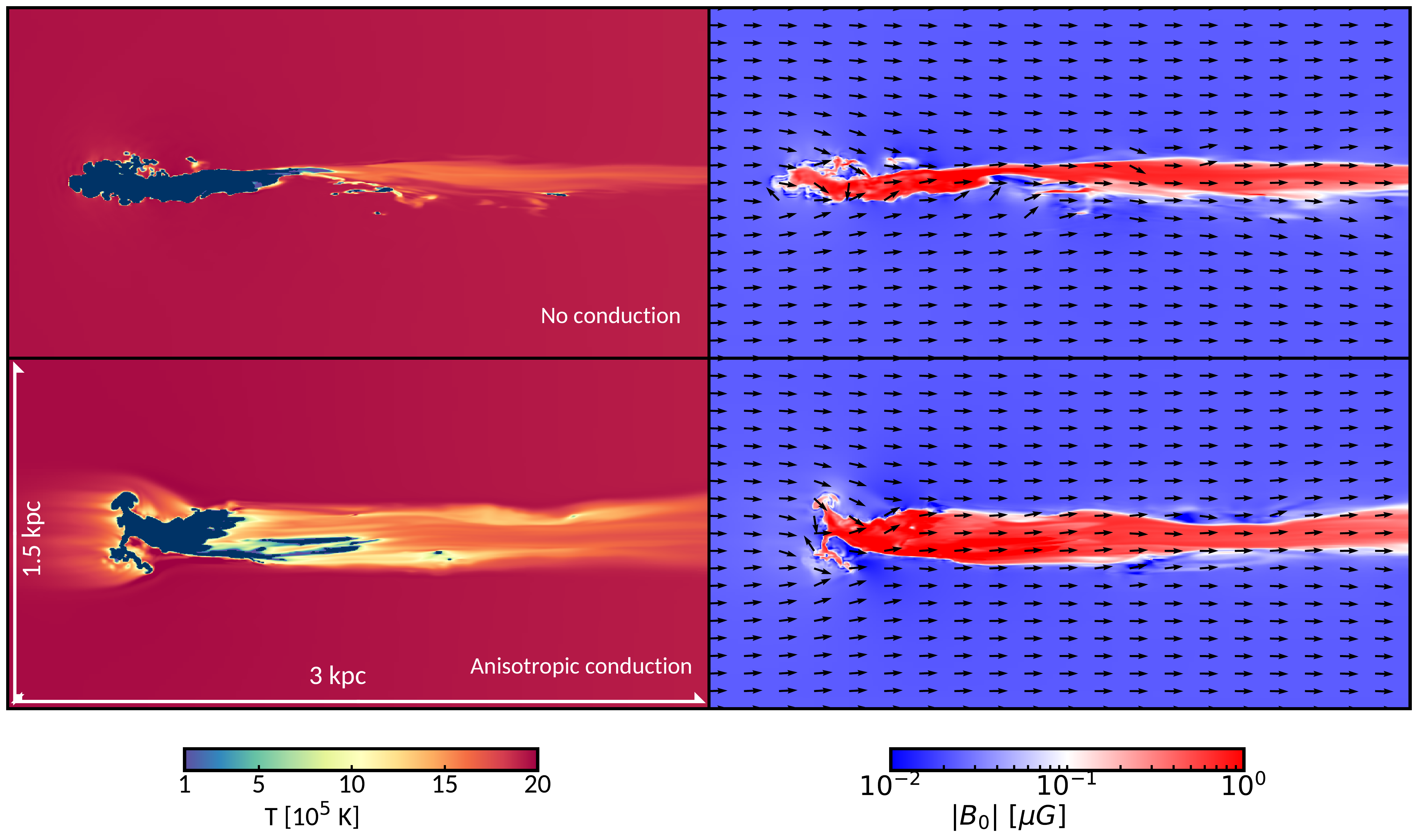}
    \caption{Slices through part of the simulation domain for simulations with a weak, parallel magnetic field without thermal conduction (top) and with anisotropic thermal conduction (bottom) for the fiducial simulation setup (Table~\ref{tab:default_parameters}). We show the temperature on the left, and the magnetic field strength (colours), and orientation (arrows) on the right. In the temperature slices we colour all gas below $2\times10^4$ K dark blue.}
    \label{fig:slice_bpar}
\end{figure*}

We show temperature and magnetic field slices with a weak field parallel to the motion of the cloud ($B_{x} = B_0$, and $B_{y}, B_{z} = 0$) in Fig.~\ref{fig:slice_bpar}. While the elongated morphology is more akin to the runs without a magnetic field, we notice that there is significantly less stripping in this case. The parallel magnetic field efficiently suppresses KH instabilities \citep{Sur_2014}, leading to a narrow wake and little stripping. In this case we notice a morphological change with the inclusion of thermal conduction. The magnetic field lines enter the cold cloud directly from the hot halo, leading to highly efficient thermal conduction at the cloud-halo interface. Similar to the HD setups with strong ($f = 0.15-0.2$) thermal conduction, this creates a temperature and density gradient at the cloud-halo interface. In this case, both the magnetic field and the temperature gradient act to suppress stripping by KH instability. 

\begin{figure*}
    \centering
    \includegraphics[width=1.0\textwidth]{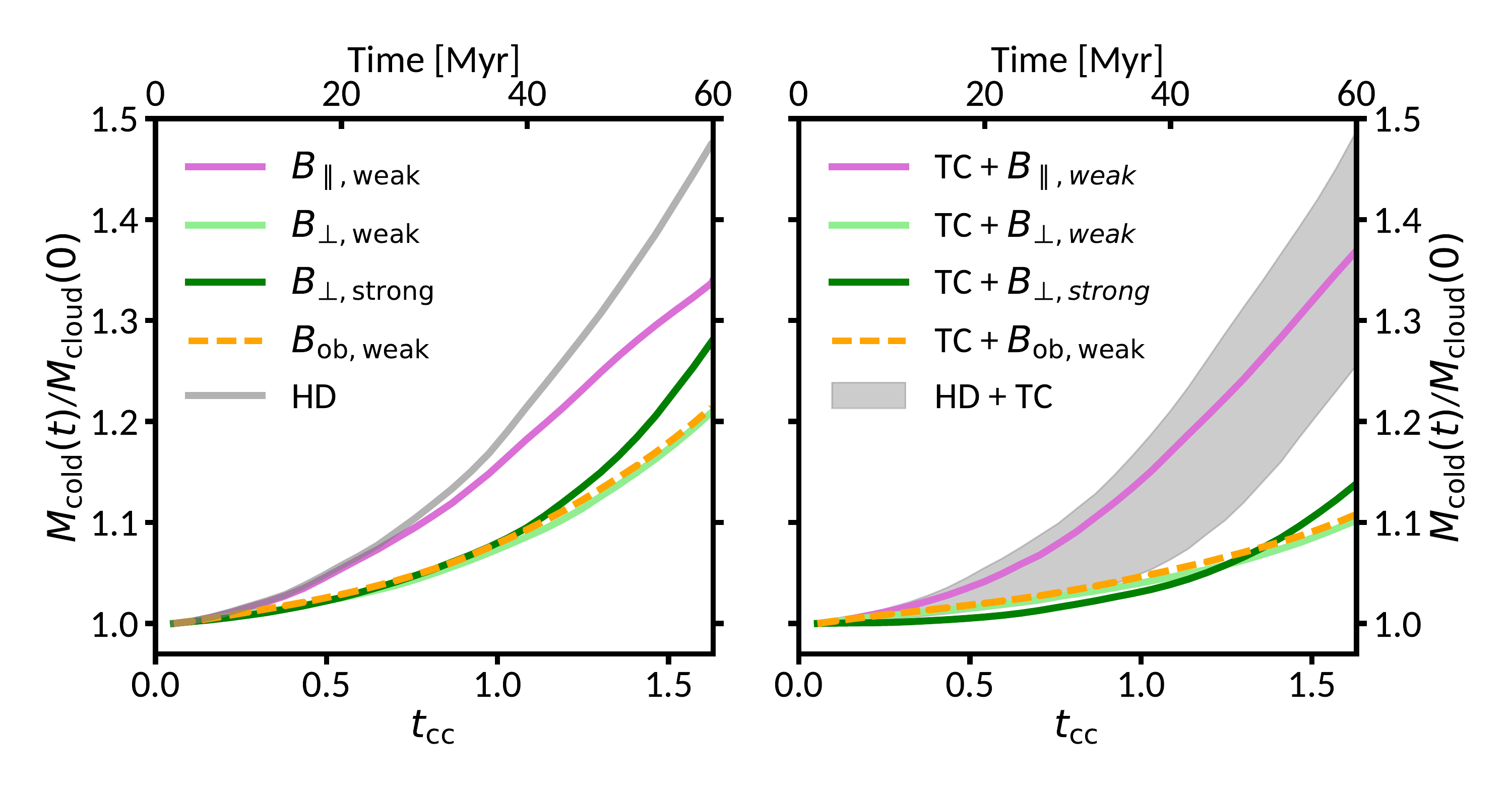}
    \caption{The cold gas evolution for our fiducial MHD simulation runs with thermal conduction turned off (left), and for runs with anisotropic thermal conduction (TC) turned on (right). The different curves refer to different strengths \& orientations of the magnetic field. Note that all simulations include radiative cooling. For comparison, we show the cold gas evolution when no magnetic field is included as seen in Fig.~\ref{fig:solar_condensation} in grey. The grey band in the right hand side plot shows the range of cold gas masses for simulations with $f=0.01$ to $f=0.2$.}
    \label{fig:mag_coldgas_default}
\end{figure*}

We show the cold gas evolution for all magnetic field setups in Fig.~\ref{fig:mag_coldgas_default}. As found previously by \citet{Gronnow_2018}, we notice that all magnetic field setups have less condensation than the HD 'only cooling' run (as shown in Fig.~\ref{fig:solar_CTC}). In the perpendicular field setups the condensation is further decreased by $\sim10\%$ when thermal conduction is included. Due to the magnetic draping effect thermal conduction does not operate efficiently on the cloud-halo interface, and stripping is mostly unhindered. Hence, the decrease in condensation is due to the evaporation of cloudlets in the wake. In contrast, for the parallel field setups the condensation changes little with the inclusion of thermal conduction. In this case the amount of gas stripped from the cloud is already very limited due to the magnetic field orientation, such that evaporation of cloudlets is negligible. The dominant way for the cloud to condense material is thus directly onto the cloud, as opposed to in the wake. The figure also shows an oblique orientation of the field ($B_{\rm ob}$, where $B_{x}= B_{y} = \frac{B_0}{\sqrt{2}}$, and $B_{z} = 0$). In this case the magnetic field is also seen to drape around the cloud, leading to very similar results as the perpendicular field setup. This shows that if magnetic draping can occur the results will be similar to the perpendicular field setup. Thus, the perpendicular magnetic field setup can be considered representative of most field orientations.

\subsubsection{The efficiency of thermal conduction}
The $f$-factor only approximates the suppression effect in the thermal conduction heat flux, so in order to isolate the effect of adding thermal conduction to the simulations we compare the ratio of cold gas masses as follows. We calculate
\begin{equation}\label{eq:delta_HD}
    \Delta M_\text{HD} = \frac{M_\text{cold, HD}}{M_\text{cold, HD + conduction}},
\end{equation}
for HD runs, and
\begin{equation}\label{eq:delta_MHD}
    \Delta M_\text{MHD} = \frac{M_\text{cold, MHD}}{M_\text{cold, MHD + conduction}},
\end{equation}
for MHD runs. In this way, the majority of the difference in the cold gas mass between HD and MHD simulations that comes from the magnetic field hindering the KH instability is filtered out. With these ratios we can thus obtain an estimate of suitable values for $f$ by comparing the approximated suppression of thermal conduction in HD simulations to the 'true' suppression in the MHD simulations. 

\begin{figure*}
    \centering
    \includegraphics[width=0.73\textwidth]{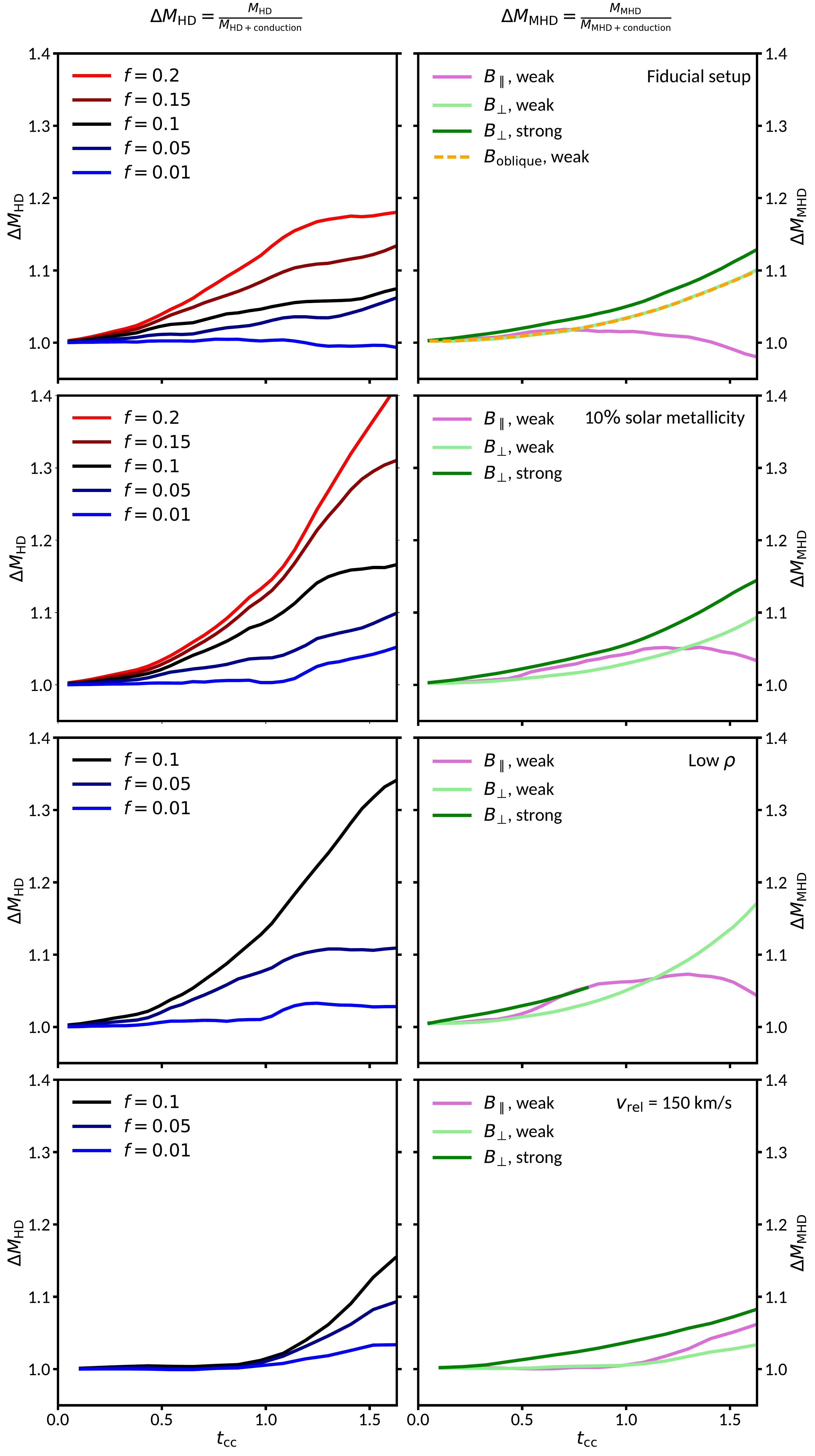}
    \caption{We show the ratio of cold gas masses with and without isotropic thermal conduction for HD simulations at a certain $f$ (left), and for MHD simulations with $f=1$, but with anisotropic thermal conduction (right). Every row represents a different simulation setup. We show from top to bottom the fiducial, 10\% solar cloud, lower density, and the high velocity setups, respectively.}
    \label{fig:all_suppressions}
\end{figure*}

In Fig.~\ref{fig:all_suppressions}, we show the results for Eqn.~\ref{eq:delta_HD} on the left, and the results for Eqn.~\ref{eq:delta_MHD} on the right, for all simulation setups. The first row contains the results for the fiducial simulation setup, seen until now. The results show percentage wise how much cold gas is formed by runs without thermal conduction, compared to runs with thermal conduction. We notice a clear distinction between field orientations, where as mentioned before, perpendicular fields are more strongly suppressed in terms of the cold gas formed than the parallel fields. For perpendicular magnetic fields we find that a thermal conduction efficiency of $f\approx0.10-0.15$ fits well, and for parallel fields the effect of thermal conduction is near negligible at $f\approx0.02$.

\subsection{Varying cloud metallicity}\label{sec:subsolar}
Typically IVC's are observed to have metallicities close to solar \citep[see e.g.][]{Richter_2001}, indicating that their origin was likely close to the metal-rich galactic disc, and were possibly ejected by the galactic fountain process \citep{Shapiro_1976, Bregman_1980, Spitoni_2008,Fraternali_2006, Fraternali_2017}. However, some IVC's \citep[see e.g.][]{Hernandez_2013, Fukui_2018} show signs of sub-solar metallicity. Here we examine the effect of lowering the cloud metallicity to 10$\%$ solar (low $Z$). This can be more typical of HVC's \citep[e.g.][]{Wakker_2007}, which we further consider in Section~\ref{sec:wind_velocity}.

\begin{figure}
    \includegraphics[width=0.5\textwidth]{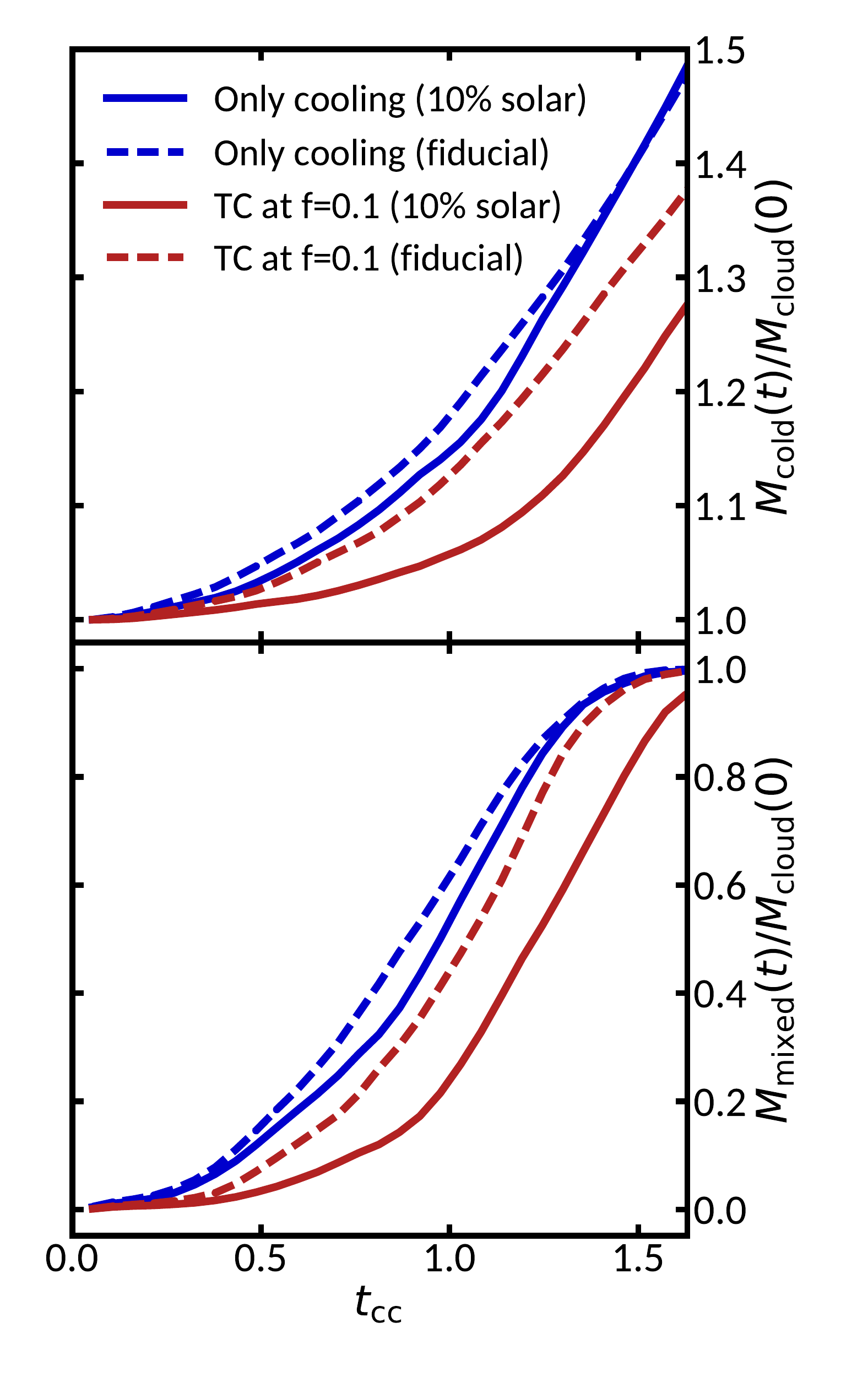}
    \caption{On the top panel we show the cold gas evolution as compared between a solar metallicity cloud (fiducial) and a 10\% solar metallicity cloud. On the bottom panel we show the mixing fraction, where the colours and markers correspond to the same simulations as in the left panel.}
    \label{fig:coldgas_metal}
\end{figure}

It is well known that gas with higher metallicity have a shorter cooling time than lower metallicity gas, due to the rapid cooling by metal lines as compared to hydrogen and helium. However, this does not necessarily lead to a significant increase in condensation in "cloud-wind" systems. An increased cooling rate has been found, for instance, to suppress the stripping of cloudlets \citep{Cooper_2009}. We also find that 
the cloudlets that are stripped and that mix with the halo gas do cool more efficiently at high $Z$ with respect to low $Z$, which in this case leads to very similar amounts of condensation, as can be seen in the left panel of Fig.~\ref{fig:coldgas_metal}. When isotropic thermal conduction is included the runs with lower cloud metallicity produce significantly less cold gas than their higher metallicity counterparts. From the definition of the Field length (Eqn.~\ref{eqn:Field}), we know that a decrease in cooling rate increases the Field length. At the peak of the cooling curve ($T\sim10^5$ K), the difference between solar and $10\%$ solar can be up to a factor 10, which increases the Field length by a factor $\sqrt{10}$. This difference in cooling rate converges around $T=10^4$ K, such that the overall change to the Field length is small, but not negligible. Thus, cloudlets are more prone to evaporate for the lower metallicity setup leading to less condensation. In addition, the mixing fraction in the right panel of Fig.~\ref{fig:coldgas_metal}, suggests that the mixing efficiency is also suppressed more strongly for a lower cloud metallicity.

\begin{figure}
\centering
    \includegraphics[width=0.5\textwidth]{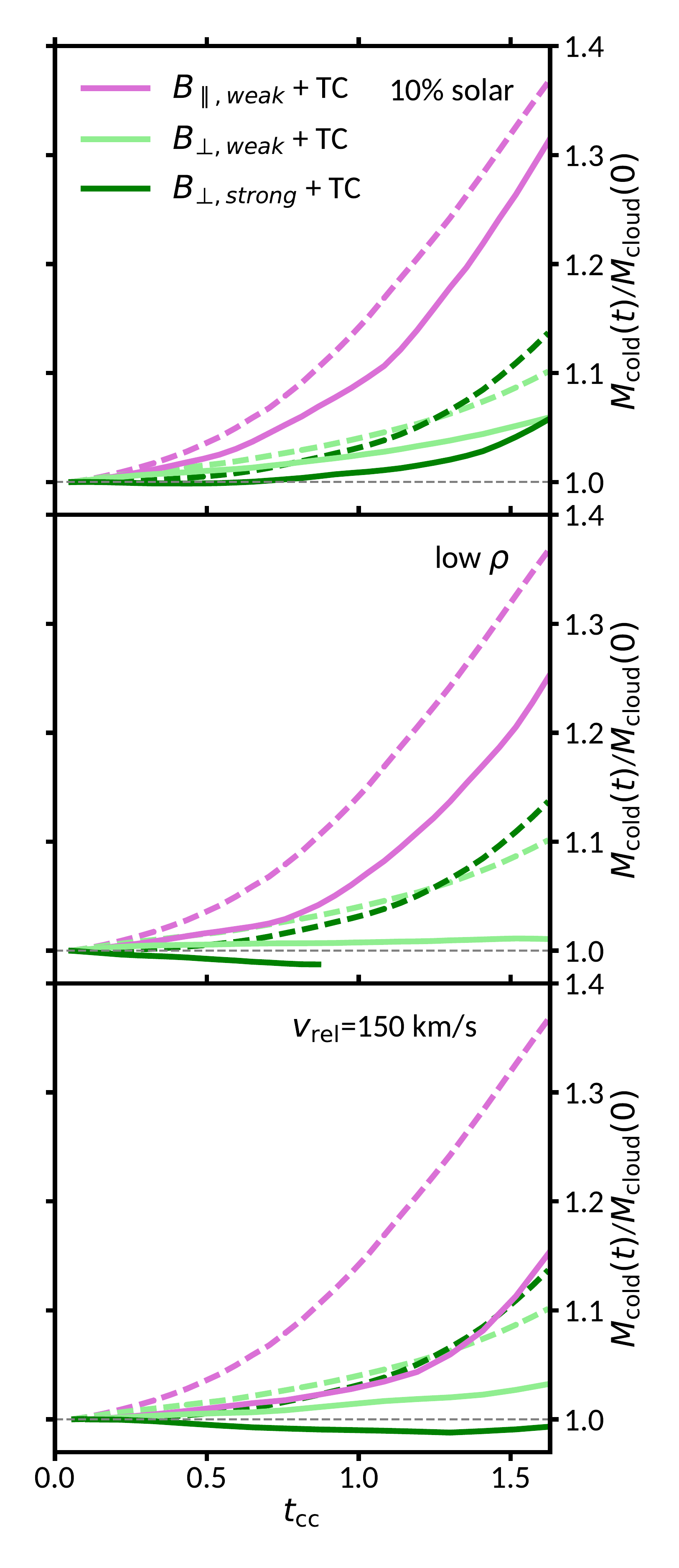}
    \caption{Cold gas evolution for the non-fiducial simulation setups including radiative cooling, magnetic fields, and anisotropic thermal conduction (TC). From left to right we show the lower cloud metallicity setup, the lower cloud/halo density setup, and the higher velocity setup, respectively. In each panel we include the fiducial simulation results as dashed lines, where the colours correspond to the same magnetic field strength/orientation.}
    \label{fig:coldgas_tc}
\end{figure}

We show the cold gas evolution for the MHD simulations with anisotropic thermal conduction in Fig.~\ref{fig:coldgas_tc} on the left panel, and the suppression effect in Fig.~\ref{fig:all_suppressions}, second row. The amount of cold gas formed is less than in the simulations with a solar metallicity cloud in all cases. The suppression effect in the MHD simulations is very similar to that in fiducial simulation setups. However, the suppression effect for the HD runs does change significantly, which suggests an overall value of $f\approx 0.05-0.10$, lower than what is found in the fiducial (high $Z$) case.

\subsection{Varying cloud and halo density}\label{sec:lowrho}
\begin{figure}
    \includegraphics[width=0.5\textwidth]{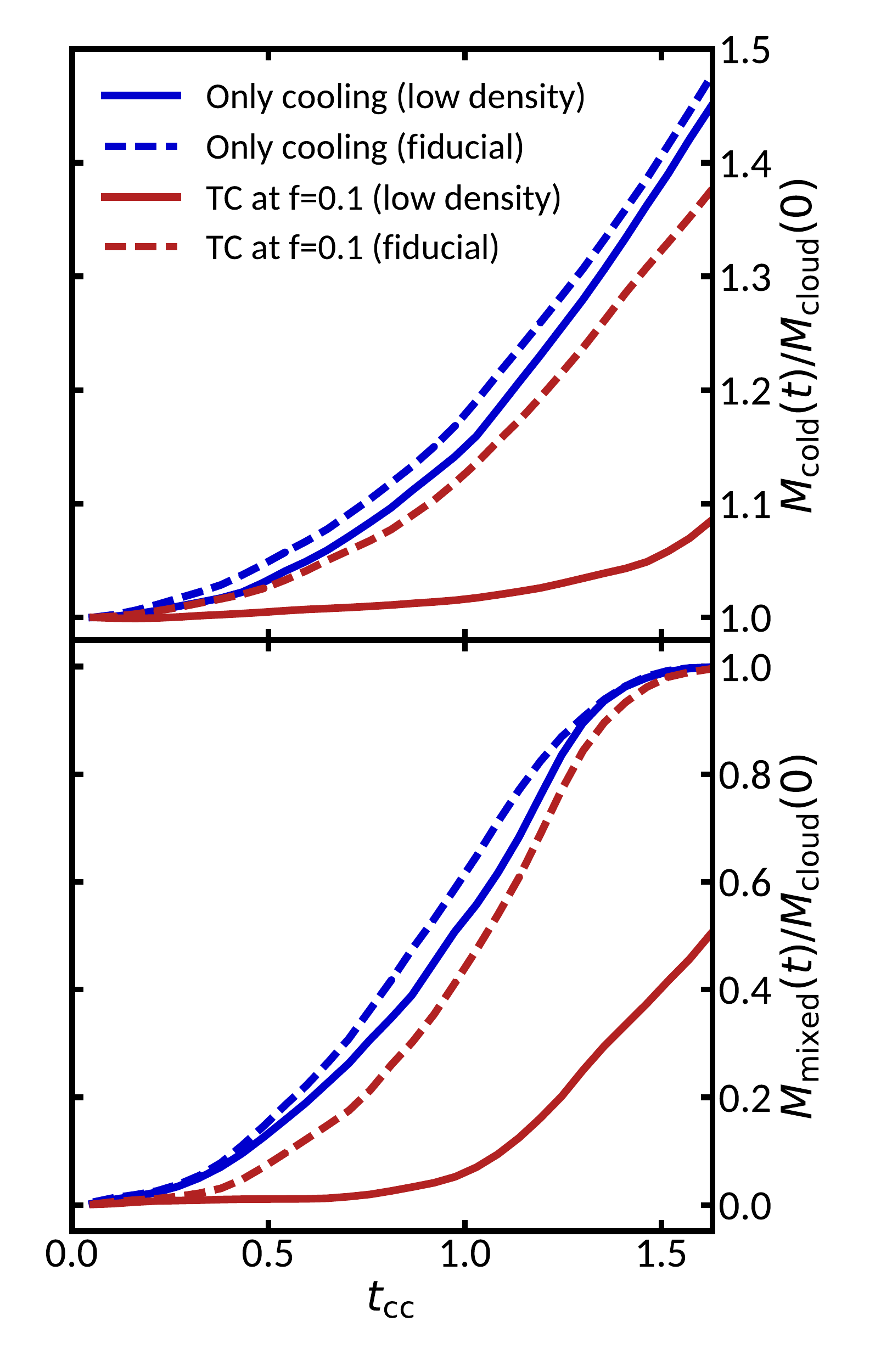}
    \caption{Cold gas evolution (top) and the mixing fraction (bottom), as compared between the fiducial simulation setup and the low-density ($n_{\rm cloud} = 0.1\ \text{cm}^{-3}$, $n_{\rm hot} = 5\times10^{-4}\ \text{cm}^{-3}$) setup.}
    \label{fig:coldgas_lowrho}
\end{figure}

We investigate a factor of 2 lower density for cloud and halo with $n_{\rm cloud}=0.1\ {\rm cm}^{-3}$ and $n_{\rm hot}=5\times10^{-4}\ {\rm cm}^{-3}$. We keep the density contrast $\chi = \frac{n_\text{cloud}}{n_\text{hot}}$ at 200, such that the pressure equilibrium between cloud and hot halo is maintained without changing their temperatures. We make the same comparison between cold gas mass and mixing rate as for the lower metallicity setup in Fig.~\ref{fig:coldgas_lowrho}. Lowering the density has little effect on the simulations with only radiative cooling. This can be understood as follows. Gas at a temperature of $10^5$ K is at the peak of the cooling curve, for which the cooling time is of the order of a few thousand years and thus immediately cools the gas to $\sim 10^4$ K. The metallicity or gas density thus does not significantly alter the evolution of the gas below $10^5$ K. Since the cooling time of gas at the halo temperature of $2\times10^6$ K is of the order of a few billion years \citep{Fraternali_2017}, the only way to obtain gas at the intermediate temperatures of $10^5$ K is to mix the cloud and halo gas. Therefore the amount of condensation is determined by the mixing efficiency. However, with the inclusion of thermal conduction a drastic decrease in condensation is seen. For this reason we have not performed simulations with $f > 0.1$, since the condensation for $f=0.1$ was already very small. For this lower density setup, the Field length increases by a factor 2. For $f> 0.1$ the Field length is of the order of the size of the cloud, which drastically alters its evolution. Besides rapid evaporation of any stripped cloudlets, there is another effect at play similar to the strong ($f=0.2$) conduction runs of the fiducial setup. Qualitative examination shows that the cloud stays mostly intact over its full evolution, with little to no stripping. The formation of KH instabilities is thus much more strongly suppressed than for the fiducial setup. The combined effect of less efficient cooling and a larger Field length is that condensation is near completely quenched for the simulation with thermal conduction at $f=0.1$. Since we see similar effects for the simulation setup with lower metallicity, it could be possible that there is a turn-off point in parameter space where thermal conduction completely inhibits condensation.  

In MHD simulations there is an additional effect as the plasma-$\beta$ parameter ($\beta=\frac{P_{\rm Th}}{P_{\rm mag}}$, where $P_{\rm Th}$ is the thermal pressure, and $P_{\rm mag}$ is the magnetic pressure) is now a factor 2 lower with respect to the fiducial setup, which means that magnetic fields are more dynamically important in this setup. The combined effect of this, and less efficient cooling is clear: almost no condensation occurs at all field strengths and orientations except a weak, parallel field, as shown in the middle panel of Fig.~\ref{fig:coldgas_tc}. We have run the strong field simulations for this setup only until $\sim0.8t_\text{cc}$, due to numerical instabilities, likely associated with the dynamically important magnetic field. Since the strong field runs have shown a consistent trend in the other simulation setups, we argue that 0.8$t_{\rm cc}$ gives a good indication of its behavior at 1.6$t_{\rm cc}$. The simulation with a strong, transverse magnetic field is not only suppressed in condensation, but shows a decrease in cold gas mass with time (evaporation). 

Similar to the 10\% solar metallicity setup, the efficiency of thermal conduction is lower as compared to the fiducial setup. In this case the suppression effect due to thermal conduction is stronger for both the HD and MHD runs (Fig.~\ref{fig:all_suppressions}, third row). We find for the lower density setup that perpendicular magnetic fields can be approximated with a thermal conduction efficiency of $f\approx0.05-0.10$. For the parallel weak field, $f\approx 0.01$.

\subsection{Varying relative velocity} \label{sec:wind_velocity}
\begin{figure}
    \includegraphics[width=0.5\textwidth]{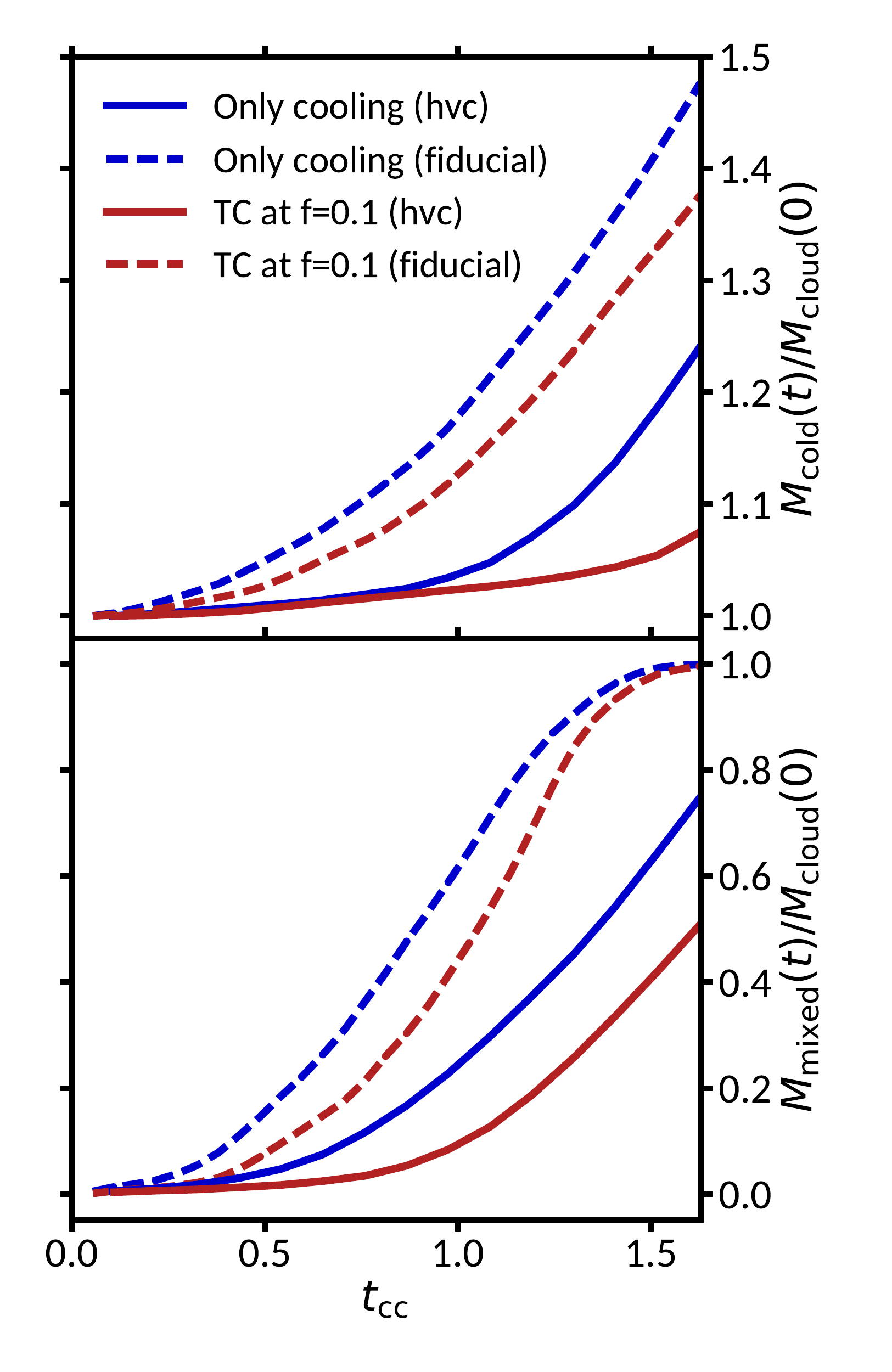}
    \caption{Cold gas mass evolution (top) and mixing fraction (bottom) for a HVC ($v_{\rm rel} = 150$ km/s) as compared to the fiducial setup. Note that $t_{\rm cc}$ is a factor 2 smaller in this HVC setup with respect to the fiducial case.}
    \label{fig:hvc_comp}
\end{figure}

In- or out-flowing gas clouds have been observed to have a wide range of velocities \citep[e.g.][]{Wakker_2007, Boomsma_2008}, and a typical distinction is made between IVC's with typical galactic fountain velocities of $\lesssim100$ km/s \citep{Fraternali_2008, Marasco_2012} with respect to the disc gas, and HVCs with higher velocities. To some extent, we can consider this distinction valid also for relative velocities between clouds and a hot halo \citep{Marinacci_2011b, Pezzulli_2016, Pezzulli_2017, Tepper-Garci_2019}. We assess the effect of the relative velocity on the evolution of our clouds by increasing it to 150 km/s ($\mathcal{M}\approx0.70$). Cloud-wind simulations with higher Mach numbers are generally harder to run numerically, but their evolution is also faster. Note that $t_\text{cc}$ is a factor 2 smaller for this HVC setup, thus these simulations are performed until $t=30$ Myrs. We show the cold gas mass evolution and mixing fractions for HD simulations in Fig.~\ref{fig:hvc_comp}. Note that similar to the lower density setup we have not performed HD simulations with $f>0.1$ since the condensation for the $f=0.1$ simulation was already very small. There is a substantial delay in the onset of condensation for the high-velocity setup. This is due to increased adiabatic heating, as shown in \citet{Gronnow_2018}. 

For the MHD simulations, we show the suppression effect in Fig.~\ref{fig:all_suppressions} on the fourth row, and the cold gas evolution in Fig.~\ref{fig:coldgas_tc} on the third row. In terms of the cloud-crushing time, we notice that there is a significant delay in the typical exponential growth rate of cold gas seen in other simulation setups \citep[see also][]{Fraternali_2015}. Even after $1.6t_\text{cc}$ the simulation with a strong, transverse magnetic field shows a net decrease in the amount of cold gas (evaporation). However, after $\approx 1.2t_\text{cc}$ this seems to increase again indicating that condensation is not indefinitely suppressed.

\section{Discussion}\label{sec:discussion}
\begin{figure*}
    \centering
    \includegraphics[width=\textwidth]{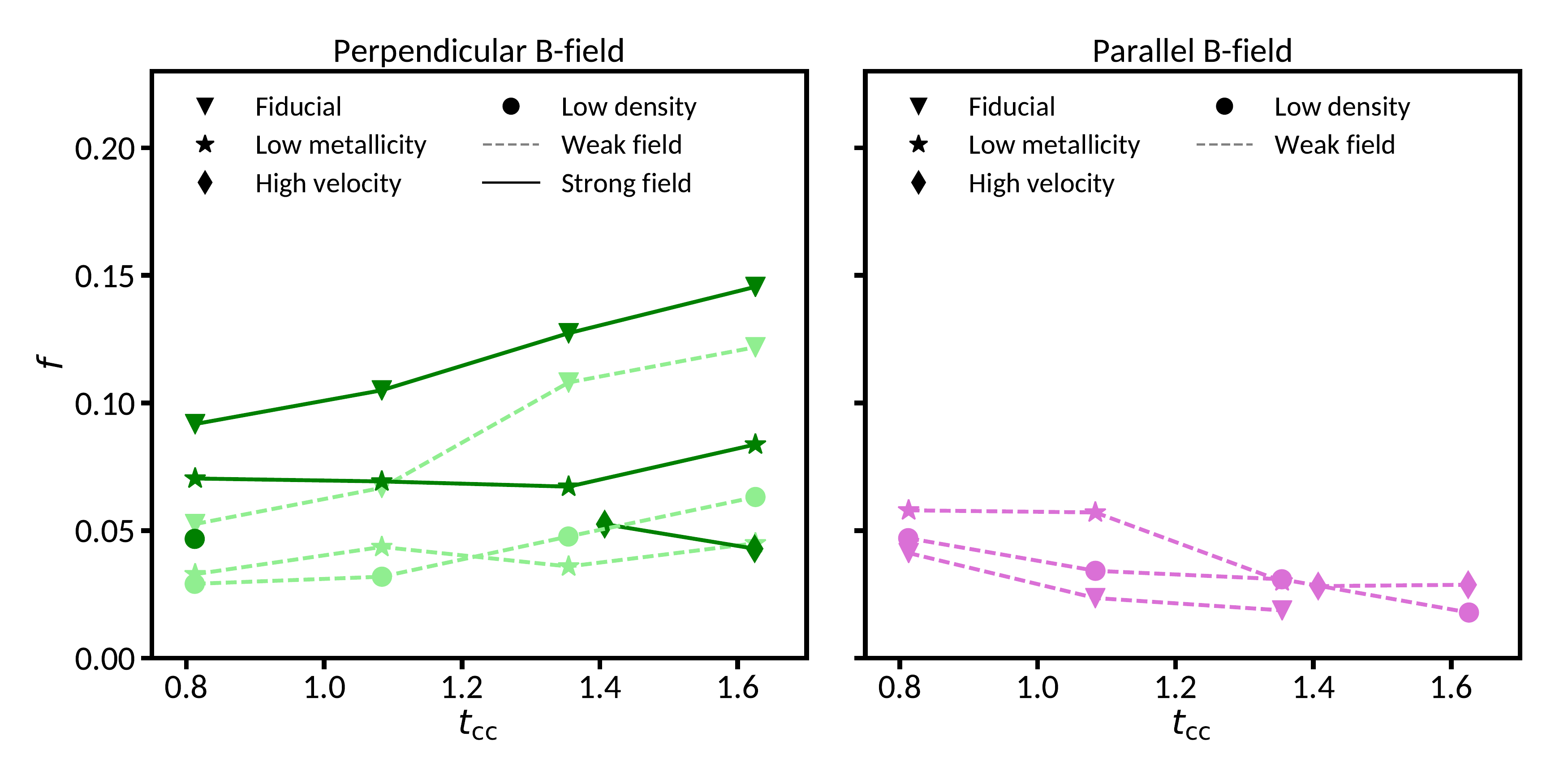}
    \caption{Linearly interpolated values for the $f$-factor sampled at regular time intervals. We do not plot the first $\sim0.8 t_\text{cc}$ since the system needs time to avoid a bias from its idealised initial conditions. There is a clear dichotomy between perpendicular and parallel magnetic fields, where thermal conduction is much less efficient in the parallel field case.}
    \label{fig:summary_plot}
\end{figure*}

We have shown and compared the evolution of HD simulations with isotropic thermal conduction to MHD simulations with anisotropic thermal conduction. In particular, we have focused on the evolution of the cold gas mass (condensation). We isolated the effect of thermal conduction on the condensation by dividing the condensation of runs without thermal conduction by the condensation of runs with thermal conduction. We then used the resulting plots (Fig.~\ref{fig:all_suppressions}) to find the HD thermal conduction efficiency $f$ by direct comparison. We now show a summary plot of $f$ as a function of time where we have compared the real suppression from the MHD results to the approximate suppression from HD results in Fig.~\ref{fig:summary_plot}. We show values after $\sim0.8t_{\rm cc}$ to allow the system to move away from its idealised initial conditions. To obtain these values we interpolate the MHD results to the curves in the HD results predicting a certain $f$. An upper limit of $f=0.15$ seems to be valid for all simulations, consistent with the upper limit of $f=0.20$ as found by \citet{Narayan_2001} for a tangled magnetic field. A clear dichotomy is also seen in different field orientations: perpendicular fields average to $0.07\pm0.04$, whereas parallel fields average to $0.03\pm0.02$, where the uncertainties are the 1$\sigma$ bounds calculated from the data points shown in Fig.~\ref{fig:summary_plot}. The difference is mainly due to the strongly suppressed stripping in the parallel field case as opposed to the perpendicular field, since thermal conduction can efficiently evaporate the stripped cloudlets. Despite the large spread in thermal conduction efficiencies, these values confirm that magnetic fields have an important effect in suppressing thermal conduction to $\lesssim10\%$ of the Spitzer value. Below we discuss the effect of resolution on our results and other potential limitations.

\subsection{Convergence tests}
We have verified for a subset of our simulations whether the results in terms of their cold gas mass evolution are converged. We perform full convergence tests for runs with only radiative cooling, radiative cooling and thermal conduction at $f=0.1$, and for radiative cooling and a weak, transverse magnetic field. We have not run the convergence test for simulations with radiative cooling, thermal conduction \textit{and} magnetic fields to 60 Myrs due to computational constraints. 

\begin{figure*}
    \centering
    \includegraphics[width=1.0\textwidth]{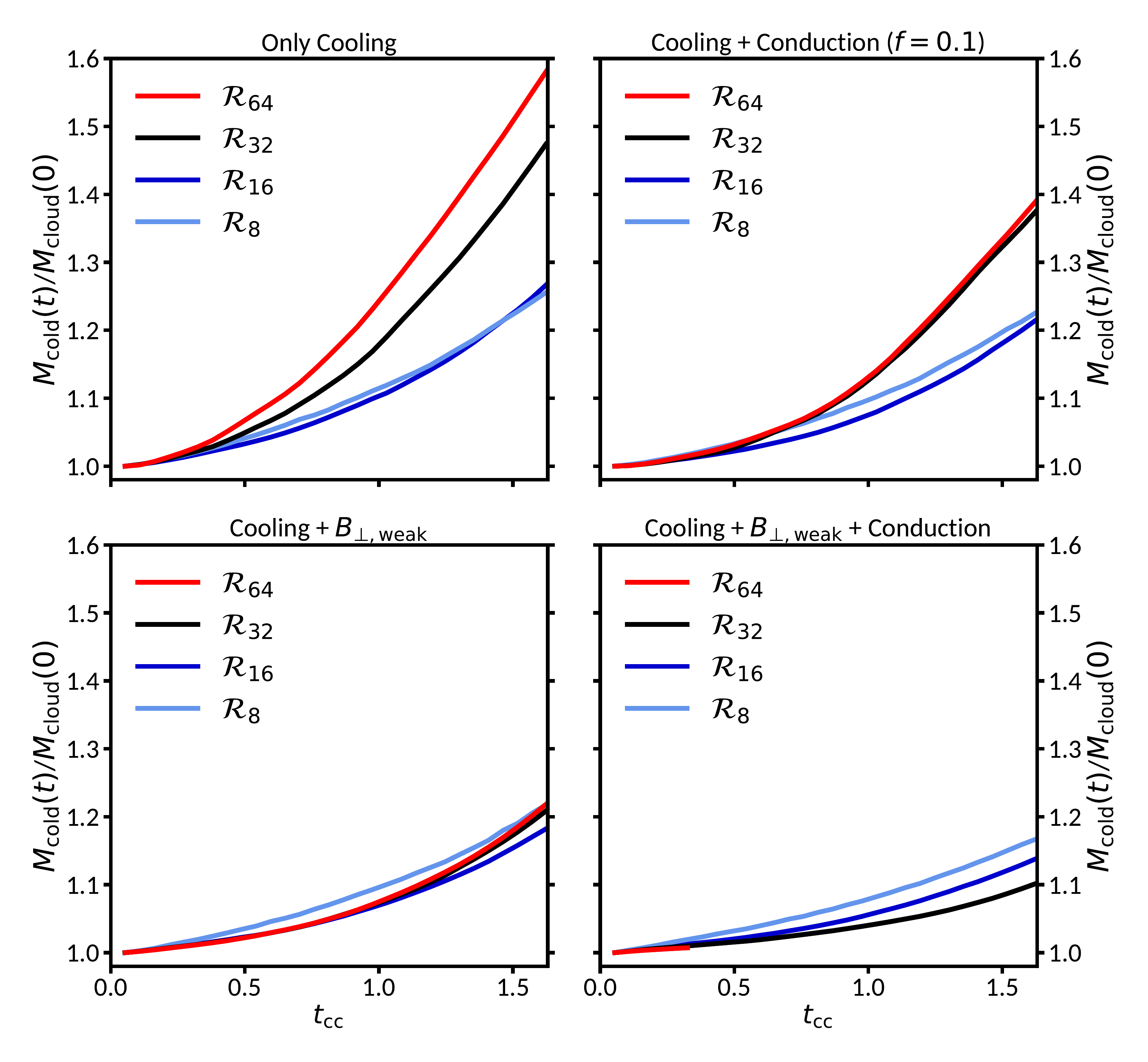}
    \caption{Convergence tests in our primary diagnostic for our fiducial simulation setup for radiative cooling (top left), radiative cooling and conduction at $f=0.1$ (top right), radiative cooling and a weak ($0.1\mu G$) perpendicular magnetic field with (right) and without (left) thermal conduction (bottom plots).}
    \label{fig:convergence}
\end{figure*}

It has been estimated that runs with only radiative cooling shatter into cold cloudlets with size $\sim0.1{\rm pc}/n$ $\approx 0.5$ pc, where $n$ is the number density of the cold material \citep{Mccourt_2018}. These sub-parsec scales are far out of our reach in terms of computational power for 3D simulations. However, thermal conduction effectively evaporates instabilities smaller than the Field length ($\lambda_{\rm F} \approx 40$ pc for $f=0.1$). For simulations with thermal conduction the resolution necessary for convergence is likely more achievable, which was also noticed by \citet[][see their Fig. 4]{Armillotta_2016}. We show our resolution studies in terms of condensation in Fig.~\ref{fig:convergence}. Note that the turbulent velocities that were added to the cloud can affect the condensation by $\sim 5\%$.
 As expected, the simulation with only radiative cooling is not converged in terms of condensation. However, with the inclusion of thermal conduction at $f=0.1$ the condensation seems to be converged to a good degree at $\mathcal{R}_{32}$. We have not performed convergence tests for runs with other values of $f$. Since our run with $f=0.1$ seems to evaporate the small scales that inhibit convergence for the cooling only run, we argue that our runs with even stronger thermal conduction with $f=0.15$ and $f=0.2$ are also converged. By the same argument, we expect that our runs with $f=0.01$ and $f=0.05$ are likely only marginally converged.  
 
  Magnetic fields can also suppress the formation of hydrodynamical instabilities \citep{Sur_2014}. \citet{Gronnow_2018} found that increasing the resolution for MHD simulations \textit{decreases} the amount of condensation, which was attributed to the width of the wake being unresolved for lower resolution. Such wider wakes could make mixing occur more efficiently. In this work, we do not see this effect. Instead, we find all resolutions predict a $\approx20\%$ increase of cold gas after $1.6t_\text{cc}$. This thin wake effect could however be obfuscated since we have added turbulent velocities to the cloud.

Finally, considering that our resolution study seems to suggest convergence for runs including thermal conduction \textit{or} a magnetic field, we might naively expect that it is likely that simulations including both thermal conduction \textit{and} magnetic fields are also converged at $\mathcal{R}_{32}$. However, the effects of anisotropic thermal conduction could affect the results differently. For this setup we have run the $\mathcal{R}_{64}$ simulation to 12 Myrs ($\approx 0.4t_{\rm cc}$), after which it was deamed computationally unfeasible. While 12 Myrs is too short to draw a reliable conclusion, we notice that the evolution is closest to that of the $\mathcal{R}_{32}$ run.

\subsection{Limitations and Future Work} \label{sec:Limitations}
While our simulations contain the typically dominant physical processes: radiative cooling, thermal conduction and magnetic fields, we have not taken into account some physics. Notably, we have not included self-gravity. Our cloud mass is $2.6\times10^4\ M_\odot$, which is much less than the Jeans mass ($M_{\rm J}\approx1.6\times10^8\ M_\odot$), so self-gravity is likely not important in this case. This conclusion is consistent with the findings of \citet{Li_2020}. Other physics not included in this work that were investigated by \citet{Li_2020} are viscosity and self-shielding of cloud material, which are also likely not important for our cloud and halo parameters. 
 Our cooling function assumes collisional ionisation equilibrium tables from \citet{Sutherland_1993}, which are slightly outdated. The collisional ionisation equilibrium assumes that the medium is optically thin for its own radiation, and that all ions are in their ground states \citep[e.g.][]{Dopita_2003}. These assumptions are very basic and in principle a chemical network based approach should be used to properly account for radiative cooling \citep[see e.g.][]{Salz_2015}. However, we do not expect significant changes to our results for different cooling functions. Furthermore, we implemented an artificial cooling floor to account for heating by a UV background. This prevents any gas from cooling to very low temperatures, but since we classify gas as cold when T $< 2\times10^4$ K the effect on the condensation will likely be negligible.
 
 We have used ordered magnetic fields for the results presented in this work. This description is idealised, since the real magnetic field in galactic halos likely has a random component \citep[e.g.][]{Klessen_2010}. However, previous work on cold clouds interacting with a wind with a random magnetic field indicates that the magnetic field still drapes around the cloud \citep{Asai_2007, Sparre_2020}. As such, the cloud will still locally experience a mostly ordered field, similar to our idealised setup. Therefore we do not expect significantly different results for setups including a random magnetic field with respect to our perpendicular field setup, which we consider to be our most realistic setup.

\section{Conclusions}\label{sec:conclusions}
Fully 3D MHD simulations including anisotropic thermal conduction of "cloud-wind" systems have not been investigated thoroughly in the literature. In this work we presented such simulations for several magnetic field strengths, orientations and parameter setups. We furthermore presented 3D HD simulations with isotropic thermal conduction at a certain efficiency $f$. Our main goal was to compare the evolution of MHD simulations with anisotropic thermal conduction to HD simulations with isotropic thermal conduction at a certain $f$.
We used the cold gas mass evolution as our primary diagnostic, and found that thermal conduction suppresses condensation in almost every case. For both HD and MHD simulations we isolated the effects of thermal conduction by dividing the cold gas mass evolution for runs with no thermal conduction to runs with thermal conduction, and directly compared the curves to find a suitable efficiency $f$.

We can draw the following conclusions:
\begin{enumerate}
    \item In HD simulations, isotropic thermal conduction always suppresses condensation, where the suppression is stronger for more efficient thermal conduction (larger $f$). This was found to be primarily due to the evaporation of small, stripped cloudlets. Condensation was found to occur most prominently close to the galactic disc ($n_{\rm hot} = 10^{-3}$ cm$^{-3}$), and was found to be substantially smaller farther away from the galaxy ($n_{\rm hot} < 5\times10^{-4}$ cm$^{-3}$). Therefore, condensation is inefficient in the outer CGM.
    
    \item In MHD simulations, anisotropic thermal conduction was found to suppress condensation more prominently for a magnetic field perpendicular to the relative velocity than for a parallel magnetic field. This was found to be due to the lack of stripped material in the parallel magnetic field setup as compared to perpendicular fields. The simulation with an oblique field produced very similar results to the perpendicular field setup, which showed that the perpendicular field setup is the most realistic.
    
    \item There is a dichotomy in efficiencies between parallel and transverse magnetic field orientations due to the distinct magnetic field interactions between the cloud and the hot medium. Transverse fields were found to have a value of $f$ in the range $0.03-0.15$, whereas parallel fields have a value of $f$ in the range $0.01-0.06$. Since perpendicular fields have been found to be the most representative field setup, an efficiency of $f$ in the range $0.03-0.15$ is likely most realistic.   
    
    \item In essentially all simulation setups there is condensation of hot halo gas. This condensation in the lower CGM environments of Milky Way-like galaxies seems unavoidable. The picture emerges that fountain accretion likely plays an important role in the recycling of halo material and can be an important regulator of the observed star formation rate in Milky Way-like galaxies.
    
\end{enumerate}
Our resolution study shows that both thermal conduction and magnetic fields can aid in reaching convergence in terms of the total amount of cold gas formed as a function of time. This is a significant result: the shattering hypothesis \citep{Mccourt_2018, Sparre_2018, Liang_2020} argues that thermal instability can shatter cold gas to very small scales. However, thermal conduction actively evaporates the smallest scales, which is likely why our run with thermal conduction appears to reach convergence. It should be noted that this is in the context of clouds moving with significant velocities with respect to the surrounding medium, and our simulations do not exclude that static or slow moving clouds in a tangled magnetic field will exhibit shattering.

We can conclude that \textit{thermal conduction plays an important role in the CGM}. In almost all simulations thermal conduction was shown to reduce condensation. Since thermal conduction is highly temperature dependent ($F \propto T^{5/2}\nabla T$), we expect that its effect on the condensation will be greater for higher temperatures. This means that galaxies with a higher virial temperature ($T>2\times10^6$ K), i.e. higher virial mass, likely experience less effective, or no fountain accretion \citep{Armillotta_2016}. 
    
While the approximation of a global efficiency of thermal conduction can in principle be used to simulate suppression due to magnetic fields, we urge caution. Magnetic fields have strong effects on cloud morphology, condensation, and in some cases the momentum transfer between cloud and halo. Therefore we stress that to obtain reliable estimates of accretion rates into galaxies and other properties linked to cloud-halo interactions in the CGM one must include magnetic fields and fully anisotropic thermal conduction.

\section*{Acknowledgements}
We thank an anonymous referee for their insightful comments and suggestions.

AG and FF acknowledge support from the Netherlands Research School for Astronomy (Nederlandse Onderzoekschool voor Astronomie, NOVA), Phase-5 research programme Network 1, Project 10.1.5.7.

The authors thank F. Marinacci, L. Armillotta for useful dicussions. We made use of \textsc{visit} \citep{VisIt}, and \textsc{paraview} \citep{Ahrens_2005} for 3D rendering. We have also made extensive use of \textsc{python} \citep{vanRossum_2009}, and \textsc{python} packages \textsc{yt} \citep{Turk_2010}, \textsc{numpy} \citep{Harris_2020}, and \textsc{matplotlib} \citep{Hunter_2007}, for the analysis and visualisation in this work.

We would furthermore like to thank the Center for Information Technology of the University of Groningen for their support and for providing access to the Peregrine high performance computing cluster. 

\section*{Data Availability}
Data available on request.



\bibliographystyle{mnras}
\bibliography{references} 




\appendix


\bsp	
\label{lastpage}
\end{document}